\documentclass[prl,twocolumn,superscriptaddress,floatfix,noshowpacs,10pt,longbibliography]{revtex4-1}%
\usepackage{graphicx,bm,times}
\usepackage{amsmath}
\usepackage{amsfonts}
\usepackage{amssymb}
\usepackage{array,multirow,textcomp}
\newcolumntype{L}[1]{>{\raggedright\let\newline\\\arraybackslash\hspace{0pt}}m{#1}}
\newcolumntype{C}[1]{>{\centering\let\newline\\\arraybackslash\hspace{0pt}}m{#1}}

\usepackage{bm}
\usepackage{color}
\usepackage{times}
\usepackage{xcolor}
\usepackage[%
  colorlinks=true,
  urlcolor=blue,
  linkcolor=blue,
  citecolor=blue
  ]{hyperref}

\begin{document}


\title{Suppression of superconductivity and enhanced critical field anisotropy in thin flakes of FeSe \\}

\author{Liam Farrar}
\email[corresponding author:]{lsf24@bath.ac.uk}
\affiliation{Centre for Nanoscience and Nanotechnology, Department of Physics, University of Bath, Bath, BA2 7AY, United Kingdom }

\author{Matthew Bristow}
\affiliation{Clarendon Laboratory, Department of Physics,
University of Oxford, Parks Road, Oxford OX1 3PU, UK}

\author{Amir A.\;Haghighirad}
\affiliation{Clarendon Laboratory, Department of Physics,
University of Oxford, Parks Road, Oxford OX1 3PU, UK}
\affiliation{Institut f\"{u}r Festk\"{o}rperphysik, Karlsruhe Institute of Technology, 76021 Karlsruhe, Germany}

\author{Alix McCollam}
\affiliation{High Field Magnet Laboratory (HFML-EMFL), Radboud University, Toernooiveld 7, 6525 ED, Nijmegen, The Netherlands}

\author{Simon J. Bending}%
\affiliation{Centre for Nanoscience and Nanotechnology, Department of Physics, University of Bath, Bath, BA2 7AY, United Kingdom }

\author{Amalia I. Coldea}
\email[corresponding author:]{amalia.coldea@physics.ox.ac.uk}
\affiliation{Clarendon Laboratory, Department of Physics,
University of Oxford, Parks Road, Oxford OX1 3PU, UK}

\date{\today}

\begin{abstract}
FeSe is a unique superconductor that can be manipulated to enhance its superconductivity using different routes
while  its monolayer form grown on different substrates reaches a record high temperature for a two-dimensional system.
In order to understand the role played by the substrate and
the reduced dimensionality on superconductivity, we examine the
superconducting properties of exfoliated FeSe thin flakes by
 reducing the thickness from bulk  down towards 9~nm.
	Magnetotransport measurements performed in magnetic fields up to 16~T and temperatures down to 2~K
help to build up complete superconducting phase diagrams of different thickness flakes.
	While the thick flakes resemble the bulk behaviour, by reducing the thickness the superconductivity of FeSe flakes is suppressed.
In the thin limit we detect signatures of a crossover towards
two-dimensional behaviour from the observation of the vortex-antivortex unbinding transition and strongly enhanced anisotropy.
Our study provides detailed insights into the evolution of the superconducting
properties from three-dimensional bulk behaviour towards the two-dimensional limit of FeSe in the absence of a dopant substrate.
\end{abstract}

\maketitle

Amongst iron-based superconductors, FeSe
 has the simplest stoichiometric crystal structure, making it an ideal candidate to
 study the mechanisms of superconductivity \cite{Hsu2008}.
Two-dimensional FeSe has attracted much interest due to the
discovery of superconductivity above 65~K in monolayer FeSe grown
on SrTiO$_{3}$ \cite{Qing-Yan2012,He2013,Ge2015a}, which is the highest critical temperature of all iron-based superconductors.
	Additionally, due to the weak van der Waals bonding of the FeSe layers,
 the material cleaves readily and has potential applications in heterostructure devices \cite{Geim2013,Novoselov}.
	It is therefore important to understand any changes in the properties of the material as it is thinned towards the monolayer limit.
	
	Previous studies examining the thickness dependence of FeSe have been limited to measurements
on thin films grown using techniques such as molecular beam expitaxy \cite{Tan2013}, pulsed laser deposition \cite{Nabeshima2013}, and DC sputtering \cite{Schneider2012}, all of which require well-optimised growth protocols.
	The resulting thin films are strongly susceptible to interaction with the growth substrate, due to factors such as strain and charge transfer.
	This can lead to effects such as the enhancement or suppression of the superconducting and structural transition temperatures when compared to
single crystals  \cite{Nabeshima2013}.
	Additionally, as the thickness of the films is reduced towards the single layer limit,
superconductivity is observed to be systematically suppressed, often resulting in a superconductor-insulator transition \cite{Wang2015}.

An alternative to the growth of thin films is to create devices by mechanical exfoliation of high quality single crystals.
	This has proven extremely successful in the case of the layered superconductors NbSe$_{2}$ \cite{Xi2016a}, TaSe$_2$ \cite{Navarro-Moratalla2016a}, and Bi$_2$Sr$_2$CaCu$_2$O$_{8+\delta}$ \cite{Jiang2014}, in which the inherent thickness-dependence
of the superconducting transition has has been measured down to a monolayer.
	Recently, exfoliated FeSe devices have been realised \cite{Lei2016,Lei2017,Ying2018}, with samples displaying superconducting behaviour at a thickness where thin films of FeSe
are typically insulating \cite{Schneider2012}. However, these samples exhibit a suppressed superconducting critical temperature when compared to the bulk crystals from which they were exfoliated. One possible factor in suppressing superconductivity is the sample degradation caused by multiple
fabrication steps, as well as long term exposure to air \cite{Yang2018b}.
	It is therefore important that any study of the thickness dependence of
superconductivity of FeSe utilise a fabrication method free from the harsh chemicals and high temperatures involved in traditional lithographic processing.

In this work we present a detailed study of the nature of superconductivity
in ultra-thin flakes of FeSe fabricated utilising a deterministic transfer method \cite{Castellanos-Gomez2014a}.
We use magnetotransport measurements in high magnetic fields up to 16~T
to investigate the effect of thickness on this materials superconducting properties.
As the thickness is reduced from 100~nm towards 9~nm, we detect a crossover towards a two-dimensional character in superconductivity
that manifests as a significant enhancement in the anisotropy of the upper critical field.
Our results give important insights into the intrinsic
nature of FeSe superconductivity in the thin limit,
unaffected by substrate interactions or other external effects.

\begin{figure}[htbp]
 \centering
  \includegraphics[trim={0cm 0cm 0cm 0cm}, width=1\linewidth,clip=true]{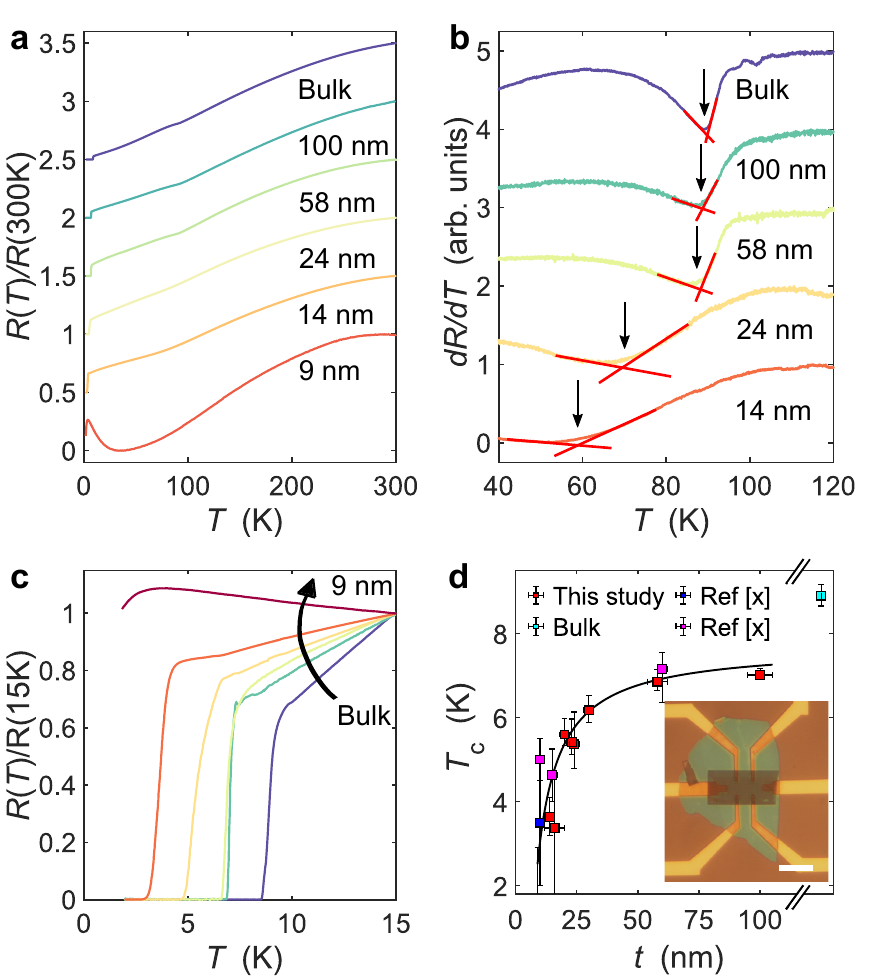}
 \caption{a) Temperature dependence of the normalised
 	resistance $R$($T$)/$R$(300~K) for a bulk crystal and five
 	different thin flakes  ($t$=9-100~nm).
 	b) The high-temperature temperature dependence of the normalised derivative of resistance, $\rm{d}\it{R}\rm{/d}\it{T}$, for the bulk sample and four thin flakes in a).
The arrows indicate the position of the structural transition at $T_{\rm{s} }$.
The curves in a) and b) are shifted vertically for clarity.
	c) The low temperature superconducting transitions
	as in a) but scaled to $R$($T$)/$R$(15~K).
	d) Thickness dependence of the superconducting critical
	temperature for several thin flake samples measured in this study, along with data from Refs.~\cite{Lei2016,Yang2018b}.
The solid line is a fit to the Cooper model \cite{Cooper1961,Simonin1986}.
	The inset shows an optical image of a 14~nm
	FeSe device capped with a thin layer ($\sim$20 nm) of h-BN.
	The scale bar corresponds to 10 $\mu$m.
}
\label{fig:Device_RT}
\end{figure}

{\bf The effect of reducing sample thickness on the transport behaviour}
	Fig.~\ref{fig:Device_RT}a) shows the typical temperature-dependence of the
normalised resistance ($R$(T)/$R$(300~K) for a bulk crystal and five thin flake devices with thickness $t$ in the range 9-100 nm
	(see Fig.~\ref{fig:RvsT_and_degradation} in Supplemental Material (SM) for additional devices).
	We observe significant changes in the transport behaviour of our devices which are highly dependent on the flake thickness.
	Firstly, the thick flake devices with $t \geq 58$ nm are of highly quality
with large RRR (see Fig.~\ref{fig:RvsT_parameters}).
They display similar transport behaviour to bulk FeSe \cite{Coldea2017}, in which the nematic phase transition occurs around $T_{\rm s} \sim$ 89~K accompanied by a superconducting transition at $T_{\rm{c}} \sim$ 9~K (see Fig. \ref{fig:Device_RT}b)).
	We notice that $T_{\rm c}$ of all thick flake devices is slightly lower than in bulk, with a maximum of $T_{\rm c} \sim$  7~K for the 100~nm flake despite a relatively high residual resistance ratio of 16 and a sharp transition width, $\Delta T_{\rm c} $ of 0.3~K.
	Next, in thinner flake devices we observe a systematic suppression of superconductivity, accompanied by a broadening of the resistive transition width, as shown in Fig. \ref{fig:Device_RT}c).
	Lastly, the thinnest device reported here with $t$ = 9~nm displays an anomalous upturn in resistance at low temperature, before a sharp decrease near 3~K, indicating that a superconducting phase may only be stabilised below the experimental temperature limit of 2~K.

	Another important signature in the transport data of FeSe is the
emergence of the nematic phase that triggers a tetragonal to
orthorhombic structural transition at $T_{\rm s}$  and causes significant in-plane distortion
	of the Fermi surface \cite{Coldea2017}.
	Fig. \ref{fig:Device_RT}b) shows that
	the nematic transition has a sharp anomaly identified by the minimum in $\rm{d}\it{R}\rm{/d}\it{T}$ that is slightly suppressed
	for thick flakes ($t>50$~nm), as compared with bulk single crystals of FeSe. 	However, in thinner samples ($t<50$ nm) this transition is ill-defined
	and appears to be significantly reduced, as shown
	in 	Fig. \ref{fig:Device_RT}b)
and Fig.~\ref{fig:RvsTSM} in SM.
	This behaviour is reminiscent of that found in polycrystalline samples of FeSe \cite{Hsu2008} or Cu-doped FeSe \cite{Zajicek2019} in which the RRR is reduced, as the degree of disorder and local inhomogeneity is much higher
	(Fig.~\ref{fig:RvsTSM} in SM),
	in comparison to high quality single crystals of FeSe in which quantum oscillations have been observed \cite{Watson}.

	A summary showing how the superconductivity of thin flakes of FeSe
	is affected by the thickness
reduction  is shown in Fig. \ref{fig:Device_RT}d).
	While $T_{\rm{c}}$  remains relatively constant for thicker flakes (50 - 100 nm),
a sharp decrease in superconductivity occurs
for the thinnest flakes ($t <25$ nm).
We can describe the observed superconducting behaviour
	 using the Cooper-law given by $T_{\rm{c}} \sim \exp(-t_m/t)$ ,
	 where $t_m = 2 a/(N_0 V)$,
	 $a$ is the Thomas Fermi screening length and
	 $N_0 V$ is the bulk pairing potential \cite{Cooper1961,Simonin1986}.
	Since $a$ is inversely proportional to the square root
	of the density of states at the Fermi energy,
	this behaviour is expected for systems
	with very small Fermi energy, as found in FeSe \cite{Coldea2017}.
	The Cooper-law is commonly used to describe superconducting thin films but the trends observed in our data are in qualitative agreement with those
 found in thin films of FeSe \cite{Schneider2012}, thin flakes of FeSe$_{0.3}$Te$_{0.7}$ \cite{Lin2015}
 and nanoflakes devices of FeSe fabricated using alternate device fabrication techniques, reported in Refs.~\cite{Lei2016,Yang2018b}.

{\bf BKT transition in thin flakes of FeSe. }
		Next, we focus on other manifestations of superconductivity in thin flake devices of FeSe.
The appearance of a Berezinskii-Kosterlitz-Thouless (BKT) transition in a material is a signature of a 2D superconducting state \cite{Berezinskit1972,Kosterlitz1973}.
	This arises due to the thermal nucleation of vortex-antivortex pairs in the absence of an external magnetic field.
	Vortex-antivortex unbinding gives rise to dissipation, which results in a resistive transition even when the temperature is below the mean field pairing temperature.
	Just above the critical current, $I_{\rm c}$, the $IV$ curves
follow the  $V \propto I^{\alpha(T)}$ dependence,
where $\alpha$ is a temperature-dependent exponent. 	
At $T_{\rm BKT}$ the critical exponent, $\alpha$, abruptly increases from 1 at higher temperatures,
due to flux flow of thermally dissociated vortex-antivortex pairs
 to 3 at lower temperatures due to the current-driven dissociation of vortex-antivortex pairs \cite{Hebard1983,Garland1987}.

\begin{figure}[htbp]
 \centering
   \includegraphics[trim={0cm 0cm 0cm 0cm}, width=1\linewidth,clip=true]{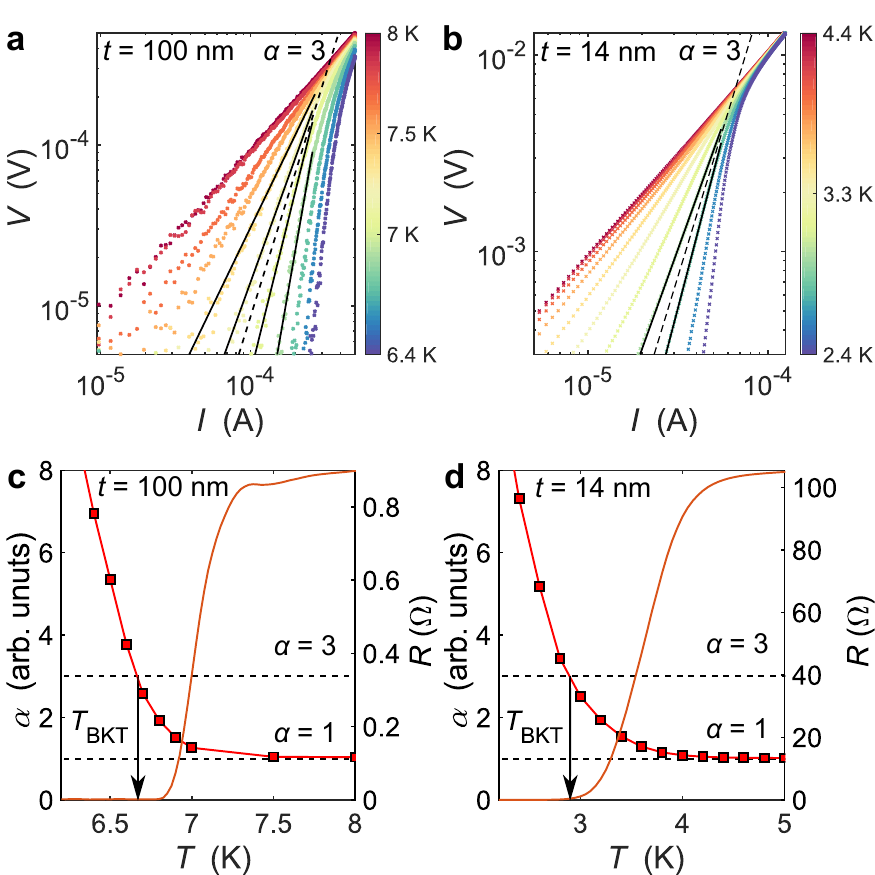}
 \caption{a,b) Voltage - current ($IV$) curves for 100~nm and 14~nm FeSe devices at  different temperatures in the vicinity of $T_{\rm c}$.
The solid black lines are linear fits to the low current regime used to extract
 the exponent $\alpha$ from the $V \propto I^{\alpha(T)}$ dependence.	
	c,d) Plot of the exponent $\alpha$ extracted from a,b) (left axis) and the resistance transition (right axis) as a function of temperature. $\alpha$($T$)
 reaches a value of 3 at $T_{\rm BKT}$ = 6.67~K for the 100~nm and $T_{\rm BKT}$ = 2.9~K  for 14~nm device.
The location of $T_{\rm BKT} $ is indicated by an arrow.
}
\label{fig:BKT}
\end{figure}

  	In order to determine whether BKT physics plays an important role in the observed suppression of superconductivity in FeSe thin flakes, we investigate the temperature-dependent current-voltage characteristics
   for two devices with $t$ =14~nm and $t$ = 100~nm at temperatures near $T_{\rm c}$, as shown in Fig. \ref{fig:BKT}a,b).
	We find a non-linear behaviour in the high current regime suggesting a current-induced vortex-antivortex depairing, as expected for a BKT transition, without displaying the sudden jump in $\alpha$($T$).
The value of the exponent
reaches the critical value of $\alpha = 3$
 at $T_{\rm BKT}$ = 2.9~K for the $t$ = 14~nm sample and $T_{\rm BKT}$ = 6.67~K for the $t$ = 100~nm sample, as shown in Fig. \ref{fig:BKT}c,d).
	In both cases, the calculated $T_{\rm BKT}$ lies below the temperature (see also Fig.~\ref{fig:BKT_transition_fit} in SM)
at which the resistance is 1$\%$ of the normal state value, suggesting that BKT physics is not the cause of the suppression of superconductivity in thin flakes.
	This appearance of a possible BKT transition is in qualitative agreement with previous reports on monolayer FeSe/SrTiO$_3$ \cite{Ying2014}, and thick films of FeSe \cite{Schneider2014}, supporting a scenario that superconductivity in FeSe is quasi-two dimensional.
The lack of the sudden jump in $\alpha $ at $T_{\rm BKT}$
and the non-linear $IV$ behavior has been found in other thin films of conventional superconductors, where the disorder smears out the sharp features \cite{Venditti2019}.
In our FeSe devices this disorder may be caused by the formation of structural domains at temperatures below the nematic transition which create local inhomogeneities, leading to an increase in the width of the superconducting transition.

\begin{figure}[t!]
\centering
   \includegraphics[trim={0cm 0cm 0cm 0cm}, width=1\linewidth,clip=true]{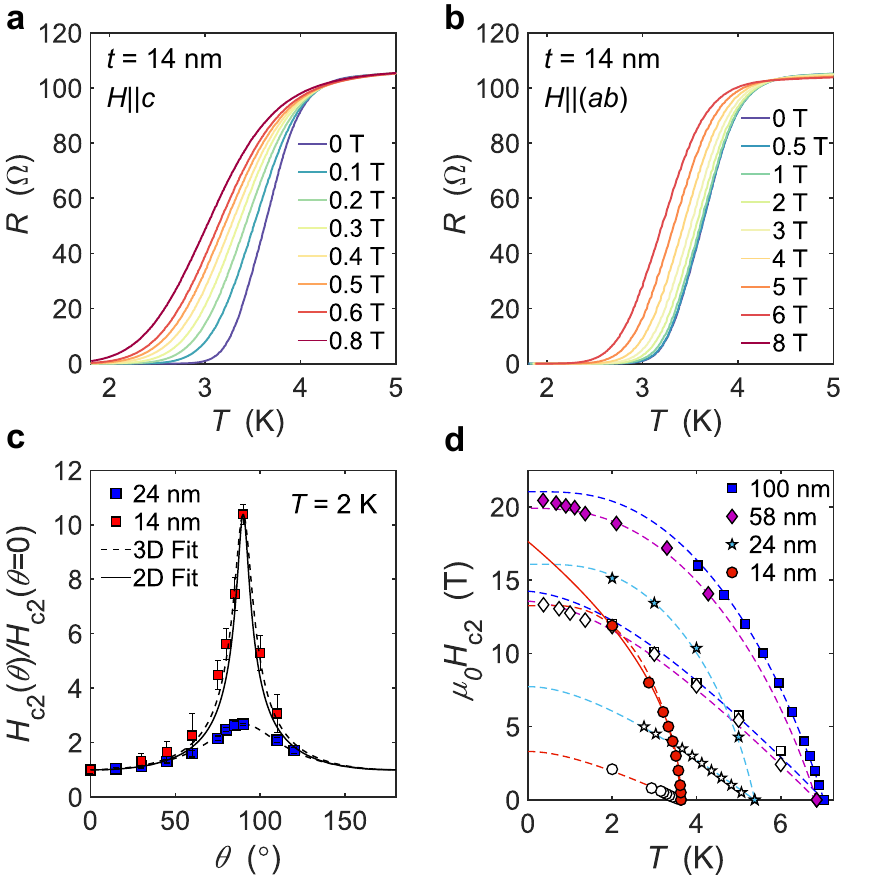}
\caption{a,b) Superconducting transition of a $t$=14~nm FeSe thin flake device as a function of magnetic field applied parallel and perpendicular to the crystallographic $c$-axis, respectively.
c) Angular dependence of the superconducting upper critical field, $H_{\rm c2}$ ($\theta$), at 2~K for devices with thickness of $t$=14~nm and $t$=24~nm
(raw data are shown in Fig.~\ref{fig:Angular_Dependence_Raw} in SM).
Dashed line represents a fit to the Ginzburg-Landau model for a type-II anisotropic superconductor and
the solid line represents a fit to the 2D-GL model \cite{Tinkham1996}.
Here $H_{\rm c2}$ is defined as the offset of superconductivity rather than the resistive transition midpoint as in the other cases.
	d) The temperature dependence of the upper critical field for four FeSe devices
with the magnetic field applied either parallel (open symbols) or perpendicular (closed symbols) to the crystallographic $c$-axis.
Dashed lines represent fits to the WHH orbital pair-breaking model and the solid line represents a fit to the 2D-GL expression for $H_{\rm c2}$ ($T$) \cite{Bennemann2008}.
}
\label{fig:anisotropy}
\end{figure}

{\bf Upper critical field in thin flakes of FeSe }
A key feature of two-dimensional superconductivity
is the significant enhancement of anisotropy in the angular dependence of the upper critical field $H_{\rm c2}$ \cite{Xi2016a}.
To investigate the suppression of superconducting behaviour in magnetic field we have measured the resistance of several devices
for different orientations of applied field.
	Figures \ref{fig:anisotropy}a,b) show the temperature dependence of the resistive transition of a thin flake device
with $t$=14~nm in constant magnetic fields up to 0.8~T with $H||c$, and in fields up to 8~T with $H||$($ab$).	
	Additional transport measurements as a function of magnetic field performed at different fixed temperatures,
as well as at fixed angles $\theta$ are shown in Fig.~\ref{fig:Angular_Dependence_Raw} in SM.
Based on these measurements, one can construct the
phase diagram of the upper critical field for several different devices, as summarised in Fig. \ref{fig:anisotropy}d).

\begin{figure}[htbp]
\centering
   \includegraphics[trim={1cm 0cm 1.5cm 0cm}, width=0.9\linewidth,clip=true]{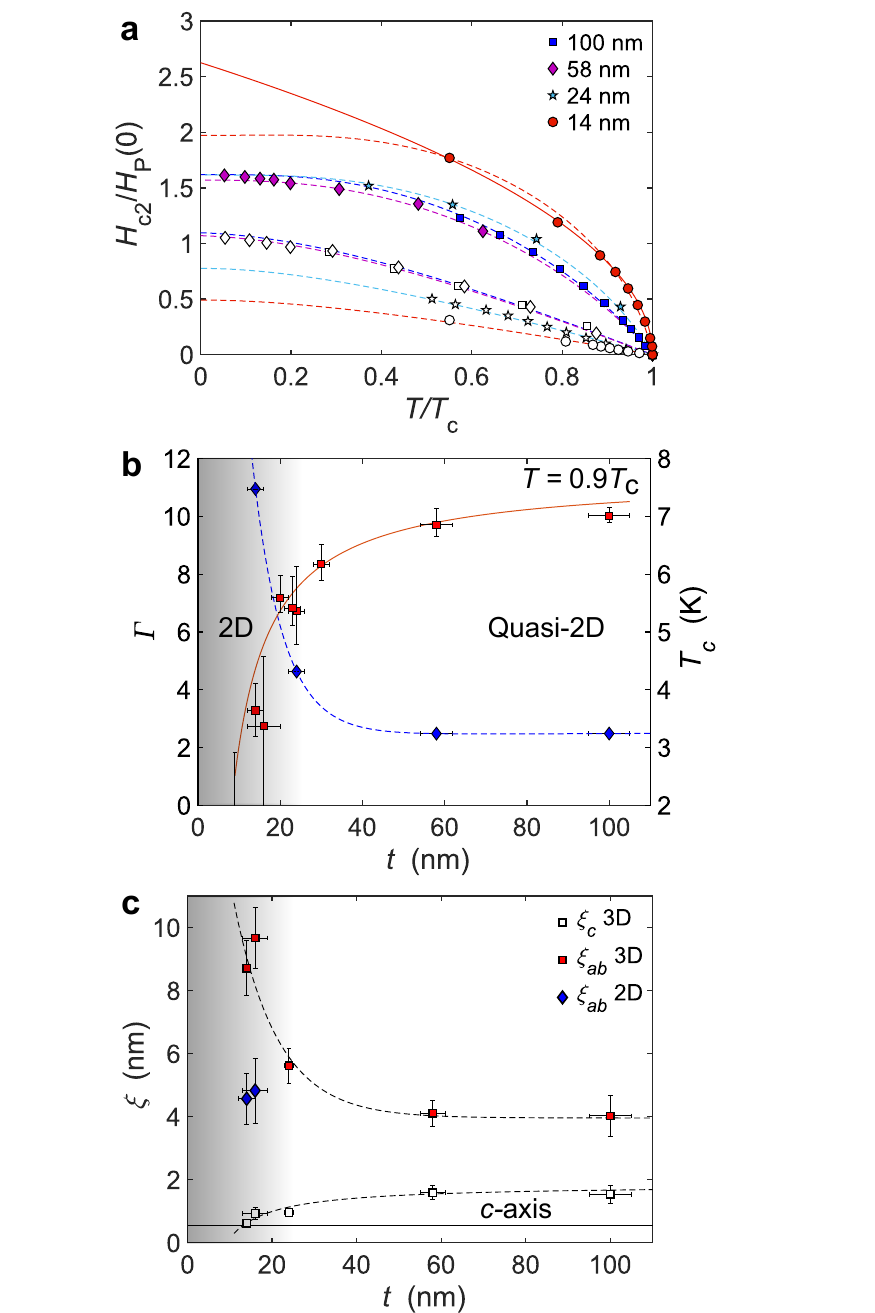}
\caption{a) The reduced superconducting critical field $H_{c2}/H_{P}$(0)
 as a function of reduced transition temperature $T/T_{\rm c}$
 for magnetic field applied either parallel (open symbols) or
 perpendicular (closed symbols) to the crystallographic $c$-axis for devices of different thickness.
Dashed lines represent fits to the WHH orbital pair-breaking model
and the solid line represents a fit to a 2D-Ginzburg Landau model (see SM) \cite{Tinkham1996}.
b) The anisotropy parameter, $\Gamma$, defined by the ratio of $H_{\rm c2}$
for the two field orientations presented in a) at $0.9T_{\rm c}$ (left axis)
and $T_{\rm c}$ (right axis) as a function of thickness, $t$.
c) The coherence lengths obtained from the slope of $H_{c2}$ in the vicinity of $T_{\rm c}$
as detailed in SM. The horizontal solid line indicates the $c$-axis layer spacing of bulk FeSe.
The shaded area indicates the crossover towards a two-dimensional highly anisotropic superconducting phase,
and the dashed lines in b) and c) are guides to the eyes.	}
\label{fig:Hc2}
\end{figure}

Additionally, to assess the changes in the superconducting anisotropy at the lowest experimental temperature
we have performed an angle-dependent study of the upper critical field
$H_{\rm c2}$($\theta$) for two devices with $t=14$~nm and $t = 24$~nm
at $T$= 2~K, as presented in Fig.~\ref{fig:anisotropy}c) using a methodology presented in Fig.~\ref{fig:Anisotropy_Comparison}.
To be able to compare the two different devices, we plot the ratio $H_{\rm c2}$($\theta$)/$H_{\rm c2}$($\theta=0)$),
as shown in Fig.~\ref{fig:Hc2}c).
Using the anisotropic Ginzburg-Landau (GL) theory \cite{Bennemann2008}, 
we can extract the anisotropy parameter
$\Gamma$ defined by the ratio between $H_{\rm c2}$ when $H ||$($ab$)
to $H_{\rm c2}$ for $H ||c$.      
We find that $\Gamma$ increases significantly from
2.4 for the $t$=24~nm device
to $>$10 when $t$=14~nm.
This indicates a significant increase in anisotropy as the flakes become thinner
and closer to the two-dimensional limit, as shown in Fig. \ref{fig:Hc2}b)
(additional data for a  $t$=16~nm flake is shown in  Fig.~\ref{fig:FeSeLF9}).
The value for the thicker flake device is comparable to the value of 1.8 observed in bulk FeSe crystals \cite{Audouard2015}, while the thinner flake device has a large anisotropy, comparable to that observed in FeS crystals \cite{Borg2016a}. This suggests that
the enhanced anisotropy can be linked to an increase in two-dimensionality of the Fermi surface.

As the superconducting anisotropy, $\Gamma$, is strongly temperature dependent,
we analyse in detail the complete superconducting phase diagram as a function of magnetic fields parallel and perpendicular to the
conducting planes for different devices, as shown in Fig. \ref{fig:anisotropy}d).
We find that the standard
three-dimensional Werthamer-Helfand-Hohenberg (WHH) model \cite{Werthamer1966},
with the inclusion of spin paramagnetism and spin-orbit scattering,
describes the temperature dependence of the upper critical field
of the thick 100~nm flake device.
A list of all obtained parameters can be found in Table~I in SM.
Orbital pair breaking alone accounts for the temperature dependence  of $H_{c2}$ for $H||c$, as shown in Fig.\ref{fig:anisotropy}d).	
However, when the magnetic field is aligned along the conducting ($ab$)
plane, a Pauli pair breaking contribution has to be included which
reduces the orbital-limited critical field by $\mu_{0} H_{\rm{P}} = \mu_{0} H_{\rm{c2}}^{\rm{orb}}/\sqrt{1 + \alpha_{\rm M}^2}$, where $\alpha_{\rm M}$ is the Maki parameter.
The extracted Maki parameter $\alpha_{\rm M}$  is 2.4 for thick flakes ($t$=54 and 100~nm),
close to the value of $\alpha_{\rm M}$ = 2.1 found for bulk single crystals \cite{Audouard2015}
For a thinner flake ($t$=24~nm), the Maki parameter increases to
4.15 (as shown in Table I in SM).
To describe the data fully the spin-orbit scattering constant
needs to be included and varies from $\lambda_{\rm SO} \sim 0.2 -0.35$ (see Fig.~\ref{fig:WHH_Fitting_Example} in SM).
Our WHH fitting parameters
are close to the values obtained for FeSe$_{0.6}$Te$_{0.6}$ single crystals
where $\alpha_{\rm M} \sim $5.5 and $\lambda_{\rm SO} \sim 1$  suggesting that
the upper critical field is dominated by Pauli paramagnetic effects \cite{Khim2010}.

In stark contrast to the behaviour
found in thick devices,
the thinnest measured device with $t$=14~nm exhibits a
drastically different temperature dependence of the upper critical field for $H_{\rm{c2}} ||$($ab$),
reaching a relatively high value of $\sim$12~T at 2~K despite the strongly suppressed $T_{\rm{c}}\sim$~3.63~K.
As a result, the slope close to $T_{\rm c}$
 increases dramatically,  predicting extremely large orbital limiting field
  $H^{\rm{orb}}_{\rm c2} = -0.69 |\textrm{d}H_{c2}/\textrm{d}T |_{T=T_{\rm{c}}} T_{\rm c}$
  (94~T as shown in Table I in SM) for  $H||$($ab$) and the Maki parameter
  becomes extremely large, $\alpha_{\rm M} \sim 11$.
This makes the WHH model less suitable to describe the experimental data
of the thinnest flakes.
Instead, we
use a 2D Ginzburg-Landau (2D-GL) theory \cite{Tinkham1996}, 
which predicts a square root temperature dependence of the in-plane $H_{c2}$  close to $T_{c}$.
This accurately describes the observed behaviour of the
$t$=14~nm device, as shown in Fig.~\ref{fig:anisotropy}d) by the solid line and in Fig.~\ref{fig:WHH_Fitting_Example} in SM. 
This finding further emphasizes
the change in character of superconductivity of FeSe in the thin limit,
becoming more two-dimensional.

{\bf Discussion}
		In order to compare the effect of thickness on the upper critical field of FeSe devices
 we investigate
 a reduced upper critical field phase diagram,
 by normalizing the upper critical field as $h=H_{\rm c2}/H_{\rm P}$(0)
 against the reduced temperature $T/T_{\rm c}$ of each device,
as shown in Fig. \ref{fig:Hc2}a). Here,
the BCS Pauli paramagnetic limit is defined
in the weak coupling limit as $H_{\rm P}$(0) = 1.85$T_{\rm{c}}$ \cite{Clogston1962}.
Interestingly, for thick devices
we observe a similar temperature dependence of the reduced upper critical field
for each orientation that can be well described by the WHH model.
Furthermore, the in-plane upper critical field
at zero temperature exceeds the Pauli limit, $H_{\rm c2}$(0) $\sim 1.6 H_{\rm P}$(0)  for thick flakes and increases above
 2 for the thinnest  $t$=14~nm flake (Fig.~\ref{fig:Hc2}a).
This suggests that spin paramagnetic effects
play an important role in determining
the  upper critical field of these thin flakes,
and indicate that FeSe thin flake may provide a possible route towards
unconventional triplet pairing \cite{Lee1997}.
Furthermore, exceeding the Pauli paramagnetic limit
coupled with a large value of the Maki parameter
without a finite $\lambda_{SO}$ would lead
to a first-order transition at low temperature, known as a FFLO state
\cite{Fulde1964,Larkin1964},
however this in not observed in our devices.
In monolayer systems such NbSe$_2$ \cite{Xi2016a} and ion gated Mos$_2$ \cite{Lu2015}
intrinsic spin-orbit interaction effects lead to Ising superconducting
and a significant increase of $H_{\rm{c2}}||${($ab$)}.
As FeSe retains inversion symmetry at all thickness,
this mechanism cannot explain the enhancement of $H_{\rm c2}${($ab$)}/$H_{\rm P}$(0)
in the $t$=14~nm device.
Moreover, in a multi-band system like FeSe
 the Pauli limit can exceed the single-band estimate since there are several gaps but a single $T_{\rm c}$ \cite{Sprau2017}
   and one could expect that the largest gap sets the Pauli limit
 \cite{Gurevich2010,Gurevich2014}.

Fig. \ref{fig:Hc2}b) summarises the thickness dependence of  the superconducting
critical temperature $T_{\rm c}$ and the upper critical field anisotropy parameter $\Gamma$ at $T$= 0.9 $T_{\rm c}$.
We observe that while $\Gamma$ increases,  the critical temperature $T_{\rm c}$ decreases suggesting the evolution
towards two-dimensional superconductivity in the thin limit of FeSe flakes. 	
To understand this
further, we use the Ginzburg-Landau formalism
to estimate the coherence lengths
from the slope of the upper critical value near $T_{\rm c}$
for the two magnetic field orientations in Fig. \ref{fig:anisotropy}d).
Fig. \ref{fig:Hc2}c) shows that
the coherence length in the ($ab$) plane, $\xi_{ab}$,
exponentially increases
from $\sim 4$~nm to  a value of $\sim 10$~nm
in the thin limit. These
 coherence lengths were estimated using the 3D GL
 and are a factor of 2 larger than the values
 extracted using 2D GL in the thin limit (see Fig. \ref{fig:Hc2}c)).
In contrast,  $\xi_{c}$
decreases significantly as the flakes get thinner,
to $\xi_{c} \sim $0.6~nm when $t=$14 nm, much smaller than the bulk
value of $\xi_{c} \sim$1-2~nm \cite{Terashima2014}, and is comparable in length to the
$c$-axis lattice constant of $\sim 0.55$~nm,
providing further evidence that
the superconductivity is becoming increasingly two-dimensional,
by  confining  the
order parameter in one unit cell of FeSe.
  In this case, the weak Josephson
 coupling of the ($ab$) planes
strongly reduces the role of orbital pair-breaking effects on $H_{\rm c2}$.
This result is somewhat surprising,
as a $t$=14~nm flake
is composed of approximately 25 individual FeSe layers,
well above the FeSe monolayer limit and not comparable in thickness to the bulk value of $\xi_{c}$.

The  superconductivity in two-dimensional superconductors
can also be suppressed by disorder.
In the 2D limit, conduction electrons
can be easily localised due to the quantum interference
effect in the presence of disorder that give rise to  Anderson localization \cite{Anderson1958}.
As the degree of disorder increases,
the superconductivity can be destroyed due to the suppression of amplitude of the superconducting order
parameter or when the phase fluctuates strongly and its coherence is lost.
	Despite a reduced $T_{\rm c}$ and lower RRR when compared to bulk, the normal state
sheet resistance of the $t=$14 nm device is $\sim 100$~$\Omega/\square$
and remains well below the quantum resistance
 ($R_{Q}=h/4e^{2}$ = 6.45 k$\Omega/\square$)
 at which a superconductor-insulator transition is expected to occur \cite{Gold1999}.
	This indicates that the suppression of $T_{\rm c}$ is not driven by disorder,
as was previously reported in amorphous thin films \cite{Schneider2014}.
In order to ensure that the observed suppression
of superconductivity and broadening effects are not extrinsic,
we have examined the effect of air exposure on $T_{\rm c}$, shown in
Fig.~\ref{fig:RvsT_and_degradation} in SM.
We find that the encapsulated FeSe thick flakes are quite robust to air exposure whereas the thinner ones are more sensitive.
However, the timescale required to significantly reduce
$T_{\rm c}$ is much longer than that
used in our study (which was less than 1 hour).

Another important parameter that can affect
the superconducting and transport behaviour of thin flakes of FeSe
is the strain induced by the substrate and its changes with temperature.
Recent work on thin films of FeSe showed that positive in-plane strain enhances $T_{\rm c}$, whilst reducing
the structural transition at $T_{\rm{s}}$ \cite{Nabeshima2013}.
This indicates that the suppression of both superconductivity
and  the structural transition in flakes
cannot be solely attributed to in-plane strain effects from the substrate.
However, the substrate inherently affects the thin flakes and can
 play a role in determining the local microstructure of the nucleated twin domain structure,
 and may lead to broader superconducting transitions in thinner samples.

The superconducting anisotropy of almost 10 detected in FeSe thin flakes
is  large compared with bulk FeSe and flakes of FeTe$_{0.55}$Se$_{0.45}$ of similar thickness.
However, large anisotropy is also found in the ultra thin limit of a
1~nm FeSe EDLT device with a large $T_{\rm c} \sim 40$~K \cite{Shiogai2018},
suggesting that the character of the two-dimensional superconductivity
is not changed by gating and doping of charge carriers.

The reduced dimensionality of thin flakes together with the
the short coherence lengths can enhance the thermal fluctuations
of the superconducting order parameter near $T_{\rm{c}}$, in comparison to
classical superconductors \cite{Blatter1994}. In thin flakes of FeSe
the type of fluctuations described by the Ginzburg number
(that can be also related to $(k_{\rm B} T_{\rm c}/E_{\rm F})^4$)
can be large due to the small Fermi energy of FeSe \cite{Coldea2017}.
This number increases upon reduction of the flake thickness, approaching
values similar to those found in cuprates \cite{Blatter1994}.
The presence of these fluctuations coupled
with the observation of the BKT transition in the thinnest flakes
supports the idea that by thinning down FeSe, one stabilises a
fluctuating two-dimensional and highly anisotropic superconductor.
The suppression of superconductivity can be either linked to strong fluctuations
or potentially to the loss of Josephson coupling between conducting layers.
As flakes become thinner, screening of the Coulomb interaction
 becomes weaker and eventually the superconductivity
is destroyed. For a system with a very small Fermi energy, such as FeSe, this mechanism
is expected to be particularly pronounced.
It remains to be understood how this type of superconductor
interacts with a substrate to drive  high-$T_{\rm c}$ superconductivity towards the single atomic layer limit.
The interface between the FeSe monolayer and the SrTiO$_3$
substrate also plays an important a role in superconductivity
due to the strain  caused by the lattice mismatch,
enhancement of electron-phonon coupling, polaronic effects
associated with the high dielectric constant of
the substrate, and carrier doping from the interface \cite{Huang2017}.

{\bf Conclusions}
In summary, we have investigated the evolution of superconductivity in high quality FeSe thin flakes devices as a function of thickness.
	We have identified a strong change in the character of superconductivity from the thick limit, in which samples show
similar behaviour to that of bulk FeSe, to the thin limit where superconductivity is strongly suppressed and highly anisotropic.
	Our studies indicate that in the thin limit and in the absence of a dopant substrate,
the superconductivity of FeSe still exhibits a two-dimensional character.
This supports the premise that enhanced two-dimensionality could be one of the key components of a high-temperature superconductor.
Future studies are needed to assess independently
the role of strain and carrier doping in stabilising the robust high-temperature superconducting state in FeSe.

{\bf Acknowledgments.}
We thank Lara Benfatto for very helpful comments on our manuscript.
The research was funded by the Oxford Centre for Applied Superconductivity (CFAS) at Oxford University.
We also acknowledge financial support of the John Fell Fund of the Oxford University.
This work was partly supported by EPSRC (EP/I004475/1, EP/I017836/1).
LF is supported by the Bath/Bristol Centre for Doctoral Training in Condensed Matter Physics, under the EPSRC (UK) Grant No. EP/L015544.
Part of this work was supported by
HFML-RU/FOM and LNCMI-CNRS, members of the European Magnetic Field
Laboratory (EMFL) and by EPSRC (UK) via its membership to the EMFL
(EP/N01085X/1)
AAH acknowledges the financial support of the Oxford Quantum
Materials Platform Grant (EP/M020517/1).
AIC acknowledges an EPSRC Career Acceleration Fellowship (EP/I004475/1).

{\bf Methods}
Thin FeSe flakes were mechanically exfoliated from high quality single crystals onto silicone elastomer polydimethylsiloxane (PDMS) stamps.
	Flakes of suitable geometry and thickness were then transferred onto Si/SiO$_{2}$ (300 nm oxide) substrates with pre-patterned Au contacts using a dry transfer set-up housed in a nitrogen glovebox with an oxygen and moisture content $<$ 1 ppm.
	To minimise environmental exposure a thin capping layer ($\sim$20 nm) of hexagonal boron nitride (h-BN) was then transferred on top of the FeSe flake, encapsulating the sample.
	An optical image of a typical sample is shown in the inset of Fig. \ref{fig:Device_RT}d).
	The thickness of each sample was accurately determined by atomic force microscopy (AFM) after all measurements had been completed.
	Magneto-transport measurements at temperatures down to 2~K and magnetic fields up to 16~T were performed using a Quantum Design Physical Property Measurement System (PPMS), with an additional sample measured at temperatures down to 0.37~K and magnetic fields up to 37.5~T at the High Field Magnet Laboratory (HFML-EMFL) in Nijmegen (shown in Fig.~\ref{fig:Hc2_Phase_Raw} in Supplemental Material (SM)).
The Hall and longitudinal resistivity contributions
were separated by (anti)symmetrizing the data
using 4-point measurements
obtained under negative and positive magnetic fields.	
	The devices presented are of high quality having a relatively high residual resistance ratio (RRR),
	$R$(300~K)/$R$(15~K) $\sim 6-16$, as detailed in Fig.~\ref{fig:RvsT_parameters} in SM.
	The superconducting critical temperature, $T_{\rm c}$,
	and upper critical field, $H_{\rm c2}$ was normally defined as the position
	 at which the resistance reached 50$\%$ of its normal state value or the maximum in its derivative.
	The upper critical field was measured
for two different orientations of the conducting
({\it ab}) plane with respect to the applied magnetic field  (either parallel to the conducting  plane, $H ||$($ab$)
($\theta$=90$^\circ$) or perpendicular to it, and
parallel to the crystallographic $c$-axis, $H || c$ ($\theta$=0$^\circ$)).
Angular-dependent studies were also performed at 2~K.

{\bf Additional Information}
Correspondence and requests for materials should
be addressed to LF (lsf24@bath.ac.uk)
and A.I.C (amalia.coldea@physics.ox.ac.uk).

\bibliographystyle{apsrev4-1}
\bibliography{PRB_Paper_July19}

\begin{thebibliography}{52}%
\makeatletter
\providecommand \@ifxundefined [1]{%
 \@ifx{#1\undefined}
}%
\providecommand \@ifnum [1]{%
 \ifnum #1\expandafter \@firstoftwo
 \else \expandafter \@secondoftwo
 \fi
}%
\providecommand \@ifx [1]{%
 \ifx #1\expandafter \@firstoftwo
 \else \expandafter \@secondoftwo
 \fi
}%
\providecommand \natexlab [1]{#1}%
\providecommand \enquote  [1]{``#1''}%
\providecommand \bibnamefont  [1]{#1}%
\providecommand \bibfnamefont [1]{#1}%
\providecommand \citenamefont [1]{#1}%
\providecommand \href@noop [0]{\@secondoftwo}%
\providecommand \href [0]{\begingroup \@sanitize@url \@href}%
\providecommand \@href[1]{\@@startlink{#1}\@@href}%
\providecommand \@@href[1]{\endgroup#1\@@endlink}%
\providecommand \@sanitize@url [0]{\catcode `\\12\catcode `\$12\catcode
  `\&12\catcode `\#12\catcode `\^12\catcode `\_12\catcode `\%12\relax}%
\providecommand \@@startlink[1]{}%
\providecommand \@@endlink[0]{}%
\providecommand \url  [0]{\begingroup\@sanitize@url \@url }%
\providecommand \@url [1]{\endgroup\@href {#1}{\urlprefix }}%
\providecommand \urlprefix  [0]{URL }%
\providecommand \Eprint [0]{\href }%
\providecommand \doibase [0]{http://dx.doi.org/}%
\providecommand \selectlanguage [0]{\@gobble}%
\providecommand \bibinfo  [0]{\@secondoftwo}%
\providecommand \bibfield  [0]{\@secondoftwo}%
\providecommand \translation [1]{[#1]}%
\providecommand \BibitemOpen [0]{}%
\providecommand \bibitemStop [0]{}%
\providecommand \bibitemNoStop [0]{.\EOS\space}%
\providecommand \EOS [0]{\spacefactor3000\relax}%
\providecommand \BibitemShut  [1]{\csname bibitem#1\endcsname}%
\let\auto@bib@innerbib\@empty
\bibitem [{\citenamefont {Hsu}\ \emph {et~al.}(2008)\citenamefont {Hsu},
  \citenamefont {Luo}, \citenamefont {Yeh}, \citenamefont {Chen}, \citenamefont
  {Huang}, \citenamefont {Wu}, \citenamefont {Lee}, \citenamefont {Huang},
  \citenamefont {Chu}, \citenamefont {Yan},\ and\ \citenamefont
  {Wu}}]{Hsu2008}%
  \BibitemOpen
  \bibfield  {author} {\bibinfo {author} {\bibfnamefont {F.-C.}\ \bibnamefont
  {Hsu}}, \bibinfo {author} {\bibfnamefont {J.-Y.}\ \bibnamefont {Luo}},
  \bibinfo {author} {\bibfnamefont {K.-W.}\ \bibnamefont {Yeh}}, \bibinfo
  {author} {\bibfnamefont {T.-K.}\ \bibnamefont {Chen}}, \bibinfo {author}
  {\bibfnamefont {T.-W.}\ \bibnamefont {Huang}}, \bibinfo {author}
  {\bibfnamefont {P.~M.}\ \bibnamefont {Wu}}, \bibinfo {author} {\bibfnamefont
  {Y.-C.}\ \bibnamefont {Lee}}, \bibinfo {author} {\bibfnamefont {Y.-L.}\
  \bibnamefont {Huang}}, \bibinfo {author} {\bibfnamefont {Y.-Y.}\ \bibnamefont
  {Chu}}, \bibinfo {author} {\bibfnamefont {D.-C.}\ \bibnamefont {Yan}}, \ and\
  \bibinfo {author} {\bibfnamefont {M.-K.}\ \bibnamefont {Wu}},\ }\href
  {\doibase 10.1073/pnas.0807325105} {\bibfield  {journal} {\bibinfo  {journal}
  {Proceedings of the National Academy of Sciences}\ }\textbf {\bibinfo
  {volume} {105}},\ \bibinfo {pages} {14262} (\bibinfo {year}
  {2008})}\BibitemShut {NoStop}%
\bibitem [{\citenamefont {Wang}\ \emph {et~al.}(2012)\citenamefont {Wang},
  \citenamefont {Li}, \citenamefont {Zhang}, \citenamefont {Zhang},
  \citenamefont {Zhang}, \citenamefont {Li}, \citenamefont {Ding},
  \citenamefont {Ou}, \citenamefont {Deng}, \citenamefont {Chang},
  \citenamefont {Wen}, \citenamefont {Song}, \citenamefont {He}, \citenamefont
  {Jia}, \citenamefont {Ji}, \citenamefont {Wang}, \citenamefont {Wang},
  \citenamefont {Chen}, \citenamefont {Ma},\ and\ \citenamefont
  {Xue}}]{Qing-Yan2012}%
  \BibitemOpen
  \bibfield  {author} {\bibinfo {author} {\bibfnamefont {Q.~Y.}\ \bibnamefont
  {Wang}}, \bibinfo {author} {\bibfnamefont {Z.}~\bibnamefont {Li}}, \bibinfo
  {author} {\bibfnamefont {W.~H.}\ \bibnamefont {Zhang}}, \bibinfo {author}
  {\bibfnamefont {Z.~C.}\ \bibnamefont {Zhang}}, \bibinfo {author}
  {\bibfnamefont {J.~S.}\ \bibnamefont {Zhang}}, \bibinfo {author}
  {\bibfnamefont {W.}~\bibnamefont {Li}}, \bibinfo {author} {\bibfnamefont
  {H.}~\bibnamefont {Ding}}, \bibinfo {author} {\bibfnamefont {Y.~B.}\
  \bibnamefont {Ou}}, \bibinfo {author} {\bibfnamefont {P.}~\bibnamefont
  {Deng}}, \bibinfo {author} {\bibfnamefont {K.}~\bibnamefont {Chang}},
  \bibinfo {author} {\bibfnamefont {J.}~\bibnamefont {Wen}}, \bibinfo {author}
  {\bibfnamefont {C.~L.}\ \bibnamefont {Song}}, \bibinfo {author}
  {\bibfnamefont {K.}~\bibnamefont {He}}, \bibinfo {author} {\bibfnamefont
  {J.~F.}\ \bibnamefont {Jia}}, \bibinfo {author} {\bibfnamefont {S.~H.}\
  \bibnamefont {Ji}}, \bibinfo {author} {\bibfnamefont {Y.~Y.}\ \bibnamefont
  {Wang}}, \bibinfo {author} {\bibfnamefont {L.~L.}\ \bibnamefont {Wang}},
  \bibinfo {author} {\bibfnamefont {X.}~\bibnamefont {Chen}}, \bibinfo {author}
  {\bibfnamefont {X.~C.}\ \bibnamefont {Ma}}, \ and\ \bibinfo {author}
  {\bibfnamefont {Q.~K.}\ \bibnamefont {Xue}},\ }\href {\doibase
  10.1088/0256-307X/29/3/037402} {\bibfield  {journal} {\bibinfo  {journal}
  {Chinese Phys. Lett.}\ }\textbf {\bibinfo {volume} {29}},\ \bibinfo {pages}
  {37402} (\bibinfo {year} {2012})}\BibitemShut {NoStop}%
\bibitem [{\citenamefont {He}\ \emph {et~al.}(2013)\citenamefont {He},
  \citenamefont {He}, \citenamefont {Zhang}, \citenamefont {Zhao},
  \citenamefont {Liu}, \citenamefont {Liu}, \citenamefont {Mou}, \citenamefont
  {Ou}, \citenamefont {Wang}, \citenamefont {Li}, \citenamefont {Wang},
  \citenamefont {Peng}, \citenamefont {Liu}, \citenamefont {Chen},
  \citenamefont {Yu}, \citenamefont {Liu}, \citenamefont {Dong}, \citenamefont
  {Zhang}, \citenamefont {Chen}, \citenamefont {Xu}, \citenamefont {Chen},
  \citenamefont {Ma}, \citenamefont {Xue},\ and\ \citenamefont
  {Zhou}}]{He2013}%
  \BibitemOpen
  \bibfield  {author} {\bibinfo {author} {\bibfnamefont {S.}~\bibnamefont
  {He}}, \bibinfo {author} {\bibfnamefont {J.}~\bibnamefont {He}}, \bibinfo
  {author} {\bibfnamefont {W.}~\bibnamefont {Zhang}}, \bibinfo {author}
  {\bibfnamefont {L.}~\bibnamefont {Zhao}}, \bibinfo {author} {\bibfnamefont
  {D.}~\bibnamefont {Liu}}, \bibinfo {author} {\bibfnamefont {X.}~\bibnamefont
  {Liu}}, \bibinfo {author} {\bibfnamefont {D.}~\bibnamefont {Mou}}, \bibinfo
  {author} {\bibfnamefont {Y.~B.}\ \bibnamefont {Ou}}, \bibinfo {author}
  {\bibfnamefont {Q.~Y.}\ \bibnamefont {Wang}}, \bibinfo {author}
  {\bibfnamefont {Z.}~\bibnamefont {Li}}, \bibinfo {author} {\bibfnamefont
  {L.}~\bibnamefont {Wang}}, \bibinfo {author} {\bibfnamefont {Y.}~\bibnamefont
  {Peng}}, \bibinfo {author} {\bibfnamefont {Y.}~\bibnamefont {Liu}}, \bibinfo
  {author} {\bibfnamefont {C.}~\bibnamefont {Chen}}, \bibinfo {author}
  {\bibfnamefont {L.}~\bibnamefont {Yu}}, \bibinfo {author} {\bibfnamefont
  {G.}~\bibnamefont {Liu}}, \bibinfo {author} {\bibfnamefont {X.}~\bibnamefont
  {Dong}}, \bibinfo {author} {\bibfnamefont {J.}~\bibnamefont {Zhang}},
  \bibinfo {author} {\bibfnamefont {C.}~\bibnamefont {Chen}}, \bibinfo {author}
  {\bibfnamefont {Z.}~\bibnamefont {Xu}}, \bibinfo {author} {\bibfnamefont
  {X.}~\bibnamefont {Chen}}, \bibinfo {author} {\bibfnamefont {X.}~\bibnamefont
  {Ma}}, \bibinfo {author} {\bibfnamefont {Q.}~\bibnamefont {Xue}}, \ and\
  \bibinfo {author} {\bibfnamefont {X.~J.}\ \bibnamefont {Zhou}},\ }\href
  {\doibase 10.1038/nmat3648} {\bibfield  {journal} {\bibinfo  {journal} {Nat.
  Mater.}\ }\textbf {\bibinfo {volume} {12}},\ \bibinfo {pages} {605} (\bibinfo
  {year} {2013})}\BibitemShut {NoStop}%
\bibitem [{\citenamefont {Ge}\ \emph {et~al.}(2015)\citenamefont {Ge},
  \citenamefont {Liu}, \citenamefont {Liu}, \citenamefont {Gao}, \citenamefont
  {Qian}, \citenamefont {Xue}, \citenamefont {Liu},\ and\ \citenamefont
  {Jia}}]{Ge2015a}%
  \BibitemOpen
  \bibfield  {author} {\bibinfo {author} {\bibfnamefont {J.~F.}\ \bibnamefont
  {Ge}}, \bibinfo {author} {\bibfnamefont {Z.~L.}\ \bibnamefont {Liu}},
  \bibinfo {author} {\bibfnamefont {C.}~\bibnamefont {Liu}}, \bibinfo {author}
  {\bibfnamefont {C.~L.}\ \bibnamefont {Gao}}, \bibinfo {author} {\bibfnamefont
  {D.}~\bibnamefont {Qian}}, \bibinfo {author} {\bibfnamefont {Q.~K.}\
  \bibnamefont {Xue}}, \bibinfo {author} {\bibfnamefont {Y.}~\bibnamefont
  {Liu}}, \ and\ \bibinfo {author} {\bibfnamefont {J.~F.}\ \bibnamefont
  {Jia}},\ }\href {\doibase 10.1038/nmat4153} {\bibfield  {journal} {\bibinfo
  {journal} {Nat. Mater.}\ }\textbf {\bibinfo {volume} {14}},\ \bibinfo {pages}
  {285} (\bibinfo {year} {2015})}\BibitemShut {NoStop}%
\bibitem [{\citenamefont {Geim}\ and\ \citenamefont
  {Grigorieva}(2013)}]{Geim2013}%
  \BibitemOpen
  \bibfield  {author} {\bibinfo {author} {\bibfnamefont {A.~K.}\ \bibnamefont
  {Geim}}\ and\ \bibinfo {author} {\bibfnamefont {I.~V.}\ \bibnamefont
  {Grigorieva}},\ }\href {\doibase 10.1038/nature12385} {\bibfield  {journal}
  {\bibinfo  {journal} {Nature}\ }\textbf {\bibinfo {volume} {499}},\ \bibinfo
  {pages} {419} (\bibinfo {year} {2013})}\BibitemShut {NoStop}%
\bibitem [{\citenamefont {Novoselov}\ \emph {et~al.}(2016)\citenamefont
  {Novoselov}, \citenamefont {Mishchenko}, \citenamefont {Carvalho},\ and\
  \citenamefont {{Castro Neto}}}]{Novoselov}%
  \BibitemOpen
  \bibfield  {author} {\bibinfo {author} {\bibfnamefont {K.~S.}\ \bibnamefont
  {Novoselov}}, \bibinfo {author} {\bibfnamefont {A.}~\bibnamefont
  {Mishchenko}}, \bibinfo {author} {\bibfnamefont {A.}~\bibnamefont
  {Carvalho}}, \ and\ \bibinfo {author} {\bibfnamefont {A.~H.}\ \bibnamefont
  {{Castro Neto}}},\ }\href {\doibase 10.1126/science.aac9439} {\bibfield
  {journal} {\bibinfo  {journal} {Science}\ }\textbf {\bibinfo {volume} {353}}
  (\bibinfo {year} {2016}),\ 10.1126/science.aac9439}\BibitemShut {NoStop}%
\bibitem [{\citenamefont {Tan}\ \emph {et~al.}(2013)\citenamefont {Tan},
  \citenamefont {Zhang}, \citenamefont {Xia}, \citenamefont {Ye}, \citenamefont
  {Chen}, \citenamefont {Xie}, \citenamefont {Peng}, \citenamefont {Xu},
  \citenamefont {Fan}, \citenamefont {Xu}, \citenamefont {Jiang}, \citenamefont
  {Zhang}, \citenamefont {Lai}, \citenamefont {Xiang}, \citenamefont {Hu},
  \citenamefont {Xie},\ and\ \citenamefont {Feng}}]{Tan2013}%
  \BibitemOpen
  \bibfield  {author} {\bibinfo {author} {\bibfnamefont {S.}~\bibnamefont
  {Tan}}, \bibinfo {author} {\bibfnamefont {Y.}~\bibnamefont {Zhang}}, \bibinfo
  {author} {\bibfnamefont {M.}~\bibnamefont {Xia}}, \bibinfo {author}
  {\bibfnamefont {Z.}~\bibnamefont {Ye}}, \bibinfo {author} {\bibfnamefont
  {F.}~\bibnamefont {Chen}}, \bibinfo {author} {\bibfnamefont {X.}~\bibnamefont
  {Xie}}, \bibinfo {author} {\bibfnamefont {R.}~\bibnamefont {Peng}}, \bibinfo
  {author} {\bibfnamefont {D.}~\bibnamefont {Xu}}, \bibinfo {author}
  {\bibfnamefont {Q.}~\bibnamefont {Fan}}, \bibinfo {author} {\bibfnamefont
  {H.}~\bibnamefont {Xu}}, \bibinfo {author} {\bibfnamefont {J.}~\bibnamefont
  {Jiang}}, \bibinfo {author} {\bibfnamefont {T.}~\bibnamefont {Zhang}},
  \bibinfo {author} {\bibfnamefont {X.}~\bibnamefont {Lai}}, \bibinfo {author}
  {\bibfnamefont {T.}~\bibnamefont {Xiang}}, \bibinfo {author} {\bibfnamefont
  {J.}~\bibnamefont {Hu}}, \bibinfo {author} {\bibfnamefont {B.}~\bibnamefont
  {Xie}}, \ and\ \bibinfo {author} {\bibfnamefont {D.}~\bibnamefont {Feng}},\
  }\href {\doibase 10.1038/nmat3654} {\bibfield  {journal} {\bibinfo  {journal}
  {Nat. Mater.}\ }\textbf {\bibinfo {volume} {12}},\ \bibinfo {pages} {634}
  (\bibinfo {year} {2013})}\BibitemShut {NoStop}%
\bibitem [{\citenamefont {Nabeshima}\ \emph {et~al.}(2013)\citenamefont
  {Nabeshima}, \citenamefont {Imai}, \citenamefont {Hanawa}, \citenamefont
  {Tsukada},\ and\ \citenamefont {Maeda}}]{Nabeshima2013}%
  \BibitemOpen
  \bibfield  {author} {\bibinfo {author} {\bibfnamefont {F.}~\bibnamefont
  {Nabeshima}}, \bibinfo {author} {\bibfnamefont {Y.}~\bibnamefont {Imai}},
  \bibinfo {author} {\bibfnamefont {M.}~\bibnamefont {Hanawa}}, \bibinfo
  {author} {\bibfnamefont {I.}~\bibnamefont {Tsukada}}, \ and\ \bibinfo
  {author} {\bibfnamefont {A.}~\bibnamefont {Maeda}},\ }\href {\doibase
  10.1063/1.4826945} {\bibfield  {journal} {\bibinfo  {journal} {Appl. Phys.
  Lett.}\ }\textbf {\bibinfo {volume} {103}},\ \bibinfo {pages} {172602}
  (\bibinfo {year} {2013})}\BibitemShut {NoStop}%
\bibitem [{\citenamefont {Schneider}\ \emph {et~al.}(2012)\citenamefont
  {Schneider}, \citenamefont {Zaitsev}, \citenamefont {Fuchs},\ and\
  \citenamefont {L{\"{o}}hneysen}}]{Schneider2012}%
  \BibitemOpen
  \bibfield  {author} {\bibinfo {author} {\bibfnamefont {R.}~\bibnamefont
  {Schneider}}, \bibinfo {author} {\bibfnamefont {A.~G.}\ \bibnamefont
  {Zaitsev}}, \bibinfo {author} {\bibfnamefont {D.}~\bibnamefont {Fuchs}}, \
  and\ \bibinfo {author} {\bibfnamefont {H.~V.}\ \bibnamefont
  {L{\"{o}}hneysen}},\ }\href {\doibase 10.1103/PhysRevLett.108.257003}
  {\bibfield  {journal} {\bibinfo  {journal} {Phys. Rev. Lett.}\ }\textbf
  {\bibinfo {volume} {108}} (\bibinfo {year} {2012}),\
  10.1103/PhysRevLett.108.257003}\BibitemShut {NoStop}%
\bibitem [{\citenamefont {Wang}\ \emph {et~al.}(2015)\citenamefont {Wang},
  \citenamefont {Zhang}, \citenamefont {Zhang}, \citenamefont {Sun},
  \citenamefont {Xing}, \citenamefont {Wang}, \citenamefont {Wang},
  \citenamefont {Ma}, \citenamefont {Xue},\ and\ \citenamefont
  {Wang}}]{Wang2015}%
  \BibitemOpen
  \bibfield  {author} {\bibinfo {author} {\bibfnamefont {Q.}~\bibnamefont
  {Wang}}, \bibinfo {author} {\bibfnamefont {W.}~\bibnamefont {Zhang}},
  \bibinfo {author} {\bibfnamefont {Z.}~\bibnamefont {Zhang}}, \bibinfo
  {author} {\bibfnamefont {Y.}~\bibnamefont {Sun}}, \bibinfo {author}
  {\bibfnamefont {Y.}~\bibnamefont {Xing}}, \bibinfo {author} {\bibfnamefont
  {Y.}~\bibnamefont {Wang}}, \bibinfo {author} {\bibfnamefont {L.}~\bibnamefont
  {Wang}}, \bibinfo {author} {\bibfnamefont {X.}~\bibnamefont {Ma}}, \bibinfo
  {author} {\bibfnamefont {Q.~K.}\ \bibnamefont {Xue}}, \ and\ \bibinfo
  {author} {\bibfnamefont {J.}~\bibnamefont {Wang}},\ }\href {\doibase
  10.1088/2053-1583/2/4/044012} {\bibfield  {journal} {\bibinfo  {journal} {2D
  Mater.}\ }\textbf {\bibinfo {volume} {2}} (\bibinfo {year} {2015}),\
  10.1088/2053-1583/2/4/044012}\BibitemShut {NoStop}%
\bibitem [{\citenamefont {Xi}\ \emph {et~al.}(2016)\citenamefont {Xi},
  \citenamefont {Wang}, \citenamefont {Zhao}, \citenamefont {Park},
  \citenamefont {Law}, \citenamefont {Berger}, \citenamefont {Forr{\'{o}}},
  \citenamefont {Shan},\ and\ \citenamefont {Mak}}]{Xi2016a}%
  \BibitemOpen
  \bibfield  {author} {\bibinfo {author} {\bibfnamefont {X.}~\bibnamefont
  {Xi}}, \bibinfo {author} {\bibfnamefont {Z.}~\bibnamefont {Wang}}, \bibinfo
  {author} {\bibfnamefont {W.}~\bibnamefont {Zhao}}, \bibinfo {author}
  {\bibfnamefont {J.~H.}\ \bibnamefont {Park}}, \bibinfo {author}
  {\bibfnamefont {K.~T.}\ \bibnamefont {Law}}, \bibinfo {author} {\bibfnamefont
  {H.}~\bibnamefont {Berger}}, \bibinfo {author} {\bibfnamefont
  {L.}~\bibnamefont {Forr{\'{o}}}}, \bibinfo {author} {\bibfnamefont
  {J.}~\bibnamefont {Shan}}, \ and\ \bibinfo {author} {\bibfnamefont {K.~F.}\
  \bibnamefont {Mak}},\ }\href {\doibase 10.1038/nphys3538} {\bibfield
  {journal} {\bibinfo  {journal} {Nat. Phys.}\ }\textbf {\bibinfo {volume}
  {12}},\ \bibinfo {pages} {139} (\bibinfo {year} {2016})}\BibitemShut
  {NoStop}%
\bibitem [{\citenamefont {Navarro-Moratalla}\ \emph {et~al.}(2016)\citenamefont
  {Navarro-Moratalla}, \citenamefont {Island}, \citenamefont
  {Man{\~{a}}s-Valero}, \citenamefont {Pinilla-Cienfuegos}, \citenamefont
  {Castellanos-Gomez}, \citenamefont {Quereda}, \citenamefont
  {Rubio-Bollinger}, \citenamefont {Chirolli}, \citenamefont
  {Silva-Guill{\'{e}}n}, \citenamefont {Agra{\"{i}}t}, \citenamefont {Steele},
  \citenamefont {Guinea}, \citenamefont {{Van Der Zant}},\ and\ \citenamefont
  {Coronado}}]{Navarro-Moratalla2016a}%
  \BibitemOpen
  \bibfield  {author} {\bibinfo {author} {\bibfnamefont {E.}~\bibnamefont
  {Navarro-Moratalla}}, \bibinfo {author} {\bibfnamefont {J.~O.}\ \bibnamefont
  {Island}}, \bibinfo {author} {\bibfnamefont {S.}~\bibnamefont
  {Man{\~{a}}s-Valero}}, \bibinfo {author} {\bibfnamefont {E.}~\bibnamefont
  {Pinilla-Cienfuegos}}, \bibinfo {author} {\bibfnamefont {A.}~\bibnamefont
  {Castellanos-Gomez}}, \bibinfo {author} {\bibfnamefont {J.}~\bibnamefont
  {Quereda}}, \bibinfo {author} {\bibfnamefont {G.}~\bibnamefont
  {Rubio-Bollinger}}, \bibinfo {author} {\bibfnamefont {L.}~\bibnamefont
  {Chirolli}}, \bibinfo {author} {\bibfnamefont {J.~A.}\ \bibnamefont
  {Silva-Guill{\'{e}}n}}, \bibinfo {author} {\bibfnamefont {N.}~\bibnamefont
  {Agra{\"{i}}t}}, \bibinfo {author} {\bibfnamefont {G.~A.}\ \bibnamefont
  {Steele}}, \bibinfo {author} {\bibfnamefont {F.}~\bibnamefont {Guinea}},
  \bibinfo {author} {\bibfnamefont {H.~S.}\ \bibnamefont {{Van Der Zant}}}, \
  and\ \bibinfo {author} {\bibfnamefont {E.}~\bibnamefont {Coronado}},\ }\href
  {\doibase 10.1038/ncomms11043} {\bibfield  {journal} {\bibinfo  {journal}
  {Nat. Commun.}\ }\textbf {\bibinfo {volume} {7}} (\bibinfo {year} {2016}),\
  10.1038/ncomms11043}\BibitemShut {NoStop}%
\bibitem [{\citenamefont {Jiang}\ \emph {et~al.}(2014)\citenamefont {Jiang},
  \citenamefont {Hu}, \citenamefont {You}, \citenamefont {Li}, \citenamefont
  {Li}, \citenamefont {Wang}, \citenamefont {Mu}, \citenamefont {Chen},
  \citenamefont {Zhang}, \citenamefont {Yu}, \citenamefont {Zhu}, \citenamefont
  {Sun}, \citenamefont {Lin}, \citenamefont {Xiao}, \citenamefont {Xie},\ and\
  \citenamefont {Jiang}}]{Jiang2014}%
  \BibitemOpen
  \bibfield  {author} {\bibinfo {author} {\bibfnamefont {D.}~\bibnamefont
  {Jiang}}, \bibinfo {author} {\bibfnamefont {T.}~\bibnamefont {Hu}}, \bibinfo
  {author} {\bibfnamefont {L.}~\bibnamefont {You}}, \bibinfo {author}
  {\bibfnamefont {Q.}~\bibnamefont {Li}}, \bibinfo {author} {\bibfnamefont
  {A.}~\bibnamefont {Li}}, \bibinfo {author} {\bibfnamefont {H.}~\bibnamefont
  {Wang}}, \bibinfo {author} {\bibfnamefont {G.}~\bibnamefont {Mu}}, \bibinfo
  {author} {\bibfnamefont {Z.}~\bibnamefont {Chen}}, \bibinfo {author}
  {\bibfnamefont {H.}~\bibnamefont {Zhang}}, \bibinfo {author} {\bibfnamefont
  {G.}~\bibnamefont {Yu}}, \bibinfo {author} {\bibfnamefont {J.}~\bibnamefont
  {Zhu}}, \bibinfo {author} {\bibfnamefont {Q.}~\bibnamefont {Sun}}, \bibinfo
  {author} {\bibfnamefont {C.}~\bibnamefont {Lin}}, \bibinfo {author}
  {\bibfnamefont {H.}~\bibnamefont {Xiao}}, \bibinfo {author} {\bibfnamefont
  {X.}~\bibnamefont {Xie}}, \ and\ \bibinfo {author} {\bibfnamefont
  {M.}~\bibnamefont {Jiang}},\ }\href {\doibase 10.1038/ncomms6708} {\bibfield
  {journal} {\bibinfo  {journal} {Nat. Commun.}\ }\textbf {\bibinfo {volume}
  {5}} (\bibinfo {year} {2014}),\ 10.1038/ncomms6708}\BibitemShut {NoStop}%
\bibitem [{\citenamefont {Lei}\ \emph {et~al.}(2016)\citenamefont {Lei},
  \citenamefont {Cui}, \citenamefont {Xiang}, \citenamefont {Shang},
  \citenamefont {Wang}, \citenamefont {Ye}, \citenamefont {Luo}, \citenamefont
  {Wu}, \citenamefont {Sun},\ and\ \citenamefont {Chen}}]{Lei2016}%
  \BibitemOpen
  \bibfield  {author} {\bibinfo {author} {\bibfnamefont {B.}~\bibnamefont
  {Lei}}, \bibinfo {author} {\bibfnamefont {J.~H.}\ \bibnamefont {Cui}},
  \bibinfo {author} {\bibfnamefont {Z.~J.}\ \bibnamefont {Xiang}}, \bibinfo
  {author} {\bibfnamefont {C.}~\bibnamefont {Shang}}, \bibinfo {author}
  {\bibfnamefont {N.~Z.}\ \bibnamefont {Wang}}, \bibinfo {author}
  {\bibfnamefont {G.~J.}\ \bibnamefont {Ye}}, \bibinfo {author} {\bibfnamefont
  {X.~G.}\ \bibnamefont {Luo}}, \bibinfo {author} {\bibfnamefont
  {T.}~\bibnamefont {Wu}}, \bibinfo {author} {\bibfnamefont {Z.}~\bibnamefont
  {Sun}}, \ and\ \bibinfo {author} {\bibfnamefont {X.~H.}\ \bibnamefont
  {Chen}},\ }\href {\doibase 10.1103/PhysRevLett.116.077002} {\bibfield
  {journal} {\bibinfo  {journal} {Phys. Rev. Lett.}\ }\textbf {\bibinfo
  {volume} {116}} (\bibinfo {year} {2016}),\
  10.1103/PhysRevLett.116.077002}\BibitemShut {NoStop}%
\bibitem [{\citenamefont {Lei}\ \emph {et~al.}(2017)\citenamefont {Lei},
  \citenamefont {Wang}, \citenamefont {Shang}, \citenamefont {Meng},
  \citenamefont {Ma}, \citenamefont {Luo}, \citenamefont {Wu}, \citenamefont
  {Sun}, \citenamefont {Wang}, \citenamefont {Jiang}, \citenamefont {Mao},
  \citenamefont {Liu}, \citenamefont {Yu}, \citenamefont {Zhang},\ and\
  \citenamefont {Chen}}]{Lei2017}%
  \BibitemOpen
  \bibfield  {author} {\bibinfo {author} {\bibfnamefont {B.}~\bibnamefont
  {Lei}}, \bibinfo {author} {\bibfnamefont {N.~Z.}\ \bibnamefont {Wang}},
  \bibinfo {author} {\bibfnamefont {C.}~\bibnamefont {Shang}}, \bibinfo
  {author} {\bibfnamefont {F.~B.}\ \bibnamefont {Meng}}, \bibinfo {author}
  {\bibfnamefont {L.~K.}\ \bibnamefont {Ma}}, \bibinfo {author} {\bibfnamefont
  {X.~G.}\ \bibnamefont {Luo}}, \bibinfo {author} {\bibfnamefont
  {T.}~\bibnamefont {Wu}}, \bibinfo {author} {\bibfnamefont {Z.}~\bibnamefont
  {Sun}}, \bibinfo {author} {\bibfnamefont {Y.}~\bibnamefont {Wang}}, \bibinfo
  {author} {\bibfnamefont {Z.}~\bibnamefont {Jiang}}, \bibinfo {author}
  {\bibfnamefont {B.~H.}\ \bibnamefont {Mao}}, \bibinfo {author} {\bibfnamefont
  {Z.}~\bibnamefont {Liu}}, \bibinfo {author} {\bibfnamefont {Y.~J.}\
  \bibnamefont {Yu}}, \bibinfo {author} {\bibfnamefont {Y.~B.}\ \bibnamefont
  {Zhang}}, \ and\ \bibinfo {author} {\bibfnamefont {X.~H.}\ \bibnamefont
  {Chen}},\ }\href {\doibase 10.1103/PhysRevB.95.020503} {\bibfield  {journal}
  {\bibinfo  {journal} {Phys. Rev. B}\ }\textbf {\bibinfo {volume} {95}},\
  \bibinfo {pages} {020503} (\bibinfo {year} {2017})}\BibitemShut {NoStop}%
\bibitem [{\citenamefont {Ying}\ \emph {et~al.}(2018)\citenamefont {Ying},
  \citenamefont {Wang}, \citenamefont {Wu}, \citenamefont {Zhao}, \citenamefont
  {Zhang}, \citenamefont {Song}, \citenamefont {Li}, \citenamefont {Lei},
  \citenamefont {Li}, \citenamefont {Yu}, \citenamefont {Cheng}, \citenamefont
  {An}, \citenamefont {Zhang}, \citenamefont {Jia}, \citenamefont {Yang},
  \citenamefont {Chen},\ and\ \citenamefont {Li}}]{Ying2018}%
  \BibitemOpen
  \bibfield  {author} {\bibinfo {author} {\bibfnamefont {T.~P.}\ \bibnamefont
  {Ying}}, \bibinfo {author} {\bibfnamefont {M.~X.}\ \bibnamefont {Wang}},
  \bibinfo {author} {\bibfnamefont {X.~X.}\ \bibnamefont {Wu}}, \bibinfo
  {author} {\bibfnamefont {Z.~Y.}\ \bibnamefont {Zhao}}, \bibinfo {author}
  {\bibfnamefont {Z.~Z.}\ \bibnamefont {Zhang}}, \bibinfo {author}
  {\bibfnamefont {B.~Q.}\ \bibnamefont {Song}}, \bibinfo {author}
  {\bibfnamefont {Y.~C.}\ \bibnamefont {Li}}, \bibinfo {author} {\bibfnamefont
  {B.}~\bibnamefont {Lei}}, \bibinfo {author} {\bibfnamefont {Q.}~\bibnamefont
  {Li}}, \bibinfo {author} {\bibfnamefont {Y.}~\bibnamefont {Yu}}, \bibinfo
  {author} {\bibfnamefont {E.~J.}\ \bibnamefont {Cheng}}, \bibinfo {author}
  {\bibfnamefont {Z.~H.}\ \bibnamefont {An}}, \bibinfo {author} {\bibfnamefont
  {Y.}~\bibnamefont {Zhang}}, \bibinfo {author} {\bibfnamefont {X.~Y.}\
  \bibnamefont {Jia}}, \bibinfo {author} {\bibfnamefont {W.}~\bibnamefont
  {Yang}}, \bibinfo {author} {\bibfnamefont {X.~H.}\ \bibnamefont {Chen}}, \
  and\ \bibinfo {author} {\bibfnamefont {S.~Y.}\ \bibnamefont {Li}},\ }\href
  {\doibase 10.1103/PhysRevLett.121.207003} {\bibfield  {journal} {\bibinfo
  {journal} {Phys. Rev. Lett.}\ }\textbf {\bibinfo {volume} {121}},\ \bibinfo
  {pages} {207003} (\bibinfo {year} {2018})}\BibitemShut {NoStop}%
\bibitem [{\citenamefont {Yang}\ \emph {et~al.}(2018)\citenamefont {Yang},
  \citenamefont {Luo}, \citenamefont {Chi}, \citenamefont {Bonn},\ and\
  \citenamefont {Xia}}]{Yang2018b}%
  \BibitemOpen
  \bibfield  {author} {\bibinfo {author} {\bibfnamefont {R.}~\bibnamefont
  {Yang}}, \bibinfo {author} {\bibfnamefont {W.}~\bibnamefont {Luo}}, \bibinfo
  {author} {\bibfnamefont {S.}~\bibnamefont {Chi}}, \bibinfo {author}
  {\bibfnamefont {D.}~\bibnamefont {Bonn}}, \ and\ \bibinfo {author}
  {\bibfnamefont {G.}~\bibnamefont {Xia}},\ }\href
  {https://arxiv.org/pdf/1805.02805.pdf} {\bibfield  {journal} {\bibinfo
  {journal} {IEEE Trans. Nanotechnol.}\ } (\bibinfo {year} {2018})}\BibitemShut
  {NoStop}%
\bibitem [{\citenamefont {Castellanos-Gomez}\ \emph {et~al.}(2014)\citenamefont
  {Castellanos-Gomez}, \citenamefont {Buscema}, \citenamefont {Molenaar},
  \citenamefont {Singh}, \citenamefont {Janssen}, \citenamefont {{Van Der
  Zant}},\ and\ \citenamefont {Steele}}]{Castellanos-Gomez2014a}%
  \BibitemOpen
  \bibfield  {author} {\bibinfo {author} {\bibfnamefont {A.}~\bibnamefont
  {Castellanos-Gomez}}, \bibinfo {author} {\bibfnamefont {M.}~\bibnamefont
  {Buscema}}, \bibinfo {author} {\bibfnamefont {R.}~\bibnamefont {Molenaar}},
  \bibinfo {author} {\bibfnamefont {V.}~\bibnamefont {Singh}}, \bibinfo
  {author} {\bibfnamefont {L.}~\bibnamefont {Janssen}}, \bibinfo {author}
  {\bibfnamefont {H.~S.}\ \bibnamefont {{Van Der Zant}}}, \ and\ \bibinfo
  {author} {\bibfnamefont {G.~A.}\ \bibnamefont {Steele}},\ }\href {\doibase
  10.1088/2053-1583/1/1/011002} {\bibfield  {journal} {\bibinfo  {journal} {2D
  Mater.}\ }\textbf {\bibinfo {volume} {1}} (\bibinfo {year} {2014}),\
  10.1088/2053-1583/1/1/011002}\BibitemShut {NoStop}%
\bibitem [{\citenamefont {Cooper}(1961)}]{Cooper1961}%
  \BibitemOpen
  \bibfield  {author} {\bibinfo {author} {\bibfnamefont {L.~N.}\ \bibnamefont
  {Cooper}},\ }\href {\doibase 10.1103/PhysRevLett.6.689} {\bibfield  {journal}
  {\bibinfo  {journal} {Phys. Rev. Lett.}\ }\textbf {\bibinfo {volume} {6}},\
  \bibinfo {pages} {689} (\bibinfo {year} {1961})}\BibitemShut {NoStop}%
\bibitem [{\citenamefont {Simonin}(1986)}]{Simonin1986}%
  \BibitemOpen
  \bibfield  {author} {\bibinfo {author} {\bibfnamefont {J.}~\bibnamefont
  {Simonin}},\ }\href {\doibase 10.1103/PhysRevB.33.7830} {\bibfield  {journal}
  {\bibinfo  {journal} {Phys. Rev. B}\ }\textbf {\bibinfo {volume} {33}},\
  \bibinfo {pages} {7830} (\bibinfo {year} {1986})}\BibitemShut {NoStop}%
\bibitem [{\citenamefont {Coldea}\ and\ \citenamefont
  {Watson}(2018)}]{Coldea2017}%
  \BibitemOpen
  \bibfield  {author} {\bibinfo {author} {\bibfnamefont {A.~I.}\ \bibnamefont
  {Coldea}}\ and\ \bibinfo {author} {\bibfnamefont {M.~D.}\ \bibnamefont
  {Watson}},\ }\href {\doibase 10.1146/annurev-conmatphys-033117-054137}
  {\bibfield  {journal} {\bibinfo  {journal} {Annu. Rev. Cond. Matt. Phys.}\
  }\textbf {\bibinfo {volume} {9}} (\bibinfo {year} {2018}),\
  10.1146/annurev-conmatphys-033117-054137}\BibitemShut {NoStop}%
\bibitem [{\citenamefont {Zajicek}\ \emph {et~al.}(2019)\citenamefont
  {Zajicek}, \citenamefont {Bristow}, \citenamefont {Haghighirad},
  \citenamefont {Singh}, \citenamefont {McCollam},\ and\ \citenamefont
  {Coldea}}]{Zajicek2019}%
  \BibitemOpen
  \bibfield  {author} {\bibinfo {author} {\bibfnamefont {Z.}~\bibnamefont
  {Zajicek}}, \bibinfo {author} {\bibfnamefont {M.}~\bibnamefont {Bristow}},
  \bibinfo {author} {\bibfnamefont {A.~A.}\ \bibnamefont {Haghighirad}},
  \bibinfo {author} {\bibfnamefont {S.}~\bibnamefont {Singh}}, \bibinfo
  {author} {\bibfnamefont {A.}~\bibnamefont {McCollam}}, \ and\ \bibinfo
  {author} {\bibfnamefont {A.~I.}\ \bibnamefont {Coldea}},\ }\href@noop {}
  {\bibfield  {journal} {\bibinfo  {journal} {in preparation}\ } (\bibinfo
  {year} {2019})}\BibitemShut {NoStop}%
\bibitem [{\citenamefont {Watson}\ \emph {et~al.}(2015)\citenamefont {Watson},
  \citenamefont {Yamashita}, \citenamefont {Kasahara}, \citenamefont {Knafo},
  \citenamefont {Nardone}, \citenamefont {B\'eard}, \citenamefont {Hardy},
  \citenamefont {McCollam}, \citenamefont {Narayanan}, \citenamefont {Blake},
  \citenamefont {Wolf}, \citenamefont {Haghighirad}, \citenamefont {Meingast},
  \citenamefont {Schofield}, \citenamefont {v.~L\"ohneysen}, \citenamefont
  {Matsuda}, \citenamefont {Coldea},\ and\ \citenamefont {Shibauchi}}]{Watson}%
  \BibitemOpen
  \bibfield  {author} {\bibinfo {author} {\bibfnamefont {M.~D.}\ \bibnamefont
  {Watson}}, \bibinfo {author} {\bibfnamefont {T.}~\bibnamefont {Yamashita}},
  \bibinfo {author} {\bibfnamefont {S.}~\bibnamefont {Kasahara}}, \bibinfo
  {author} {\bibfnamefont {W.}~\bibnamefont {Knafo}}, \bibinfo {author}
  {\bibfnamefont {M.}~\bibnamefont {Nardone}}, \bibinfo {author} {\bibfnamefont
  {J.}~\bibnamefont {B\'eard}}, \bibinfo {author} {\bibfnamefont
  {F.}~\bibnamefont {Hardy}}, \bibinfo {author} {\bibfnamefont
  {A.}~\bibnamefont {McCollam}}, \bibinfo {author} {\bibfnamefont
  {A.}~\bibnamefont {Narayanan}}, \bibinfo {author} {\bibfnamefont {S.~F.}\
  \bibnamefont {Blake}}, \bibinfo {author} {\bibfnamefont {T.}~\bibnamefont
  {Wolf}}, \bibinfo {author} {\bibfnamefont {A.~A.}\ \bibnamefont
  {Haghighirad}}, \bibinfo {author} {\bibfnamefont {C.}~\bibnamefont
  {Meingast}}, \bibinfo {author} {\bibfnamefont {A.~J.}\ \bibnamefont
  {Schofield}}, \bibinfo {author} {\bibfnamefont {H.}~\bibnamefont
  {v.~L\"ohneysen}}, \bibinfo {author} {\bibfnamefont {Y.}~\bibnamefont
  {Matsuda}}, \bibinfo {author} {\bibfnamefont {A.~I.}\ \bibnamefont {Coldea}},
  \ and\ \bibinfo {author} {\bibfnamefont {T.}~\bibnamefont {Shibauchi}},\
  }\href {\doibase 10.1103/PhysRevLett.115.027006} {\bibfield  {journal}
  {\bibinfo  {journal} {Phys. Rev. Lett.}\ }\textbf {\bibinfo {volume} {115}},\
  \bibinfo {pages} {027006} (\bibinfo {year} {2015})}\BibitemShut {NoStop}%
\bibitem [{\citenamefont {Lin}\ \emph {et~al.}(2015)\citenamefont {Lin},
  \citenamefont {Mei}, \citenamefont {Wei}, \citenamefont {Sun}, \citenamefont
  {Wu}, \citenamefont {Huang}, \citenamefont {Zhang}, \citenamefont {Liu},
  \citenamefont {Feng}, \citenamefont {Tian}, \citenamefont {Yang},
  \citenamefont {Li}, \citenamefont {Wang}, \citenamefont {Zhang},
  \citenamefont {Lu},\ and\ \citenamefont {Zhao}}]{Lin2015}%
  \BibitemOpen
  \bibfield  {author} {\bibinfo {author} {\bibfnamefont {Z.}~\bibnamefont
  {Lin}}, \bibinfo {author} {\bibfnamefont {C.}~\bibnamefont {Mei}}, \bibinfo
  {author} {\bibfnamefont {L.}~\bibnamefont {Wei}}, \bibinfo {author}
  {\bibfnamefont {Z.}~\bibnamefont {Sun}}, \bibinfo {author} {\bibfnamefont
  {S.}~\bibnamefont {Wu}}, \bibinfo {author} {\bibfnamefont {H.}~\bibnamefont
  {Huang}}, \bibinfo {author} {\bibfnamefont {S.}~\bibnamefont {Zhang}},
  \bibinfo {author} {\bibfnamefont {C.}~\bibnamefont {Liu}}, \bibinfo {author}
  {\bibfnamefont {Y.}~\bibnamefont {Feng}}, \bibinfo {author} {\bibfnamefont
  {H.}~\bibnamefont {Tian}}, \bibinfo {author} {\bibfnamefont {H.}~\bibnamefont
  {Yang}}, \bibinfo {author} {\bibfnamefont {J.}~\bibnamefont {Li}}, \bibinfo
  {author} {\bibfnamefont {Y.}~\bibnamefont {Wang}}, \bibinfo {author}
  {\bibfnamefont {G.}~\bibnamefont {Zhang}}, \bibinfo {author} {\bibfnamefont
  {Y.}~\bibnamefont {Lu}}, \ and\ \bibinfo {author} {\bibfnamefont
  {Y.}~\bibnamefont {Zhao}},\ }\href {\doibase 10.1038/srep14133} {\bibfield
  {journal} {\bibinfo  {journal} {Scientific Reports}\ }\textbf {\bibinfo
  {volume} {5}},\ \bibinfo {pages} {14133} (\bibinfo {year}
  {2015})}\BibitemShut {NoStop}%
\bibitem [{\citenamefont {Berezinskii}(1971)}]{Berezinskit1972}%
  \BibitemOpen
  \bibfield  {author} {\bibinfo {author} {\bibfnamefont {V.~.~L.}\ \bibnamefont
  {Berezinskii}},\ }\href {\doibase 1971JETP...32..493B} {\bibfield  {journal}
  {\bibinfo  {journal} {J. Exp. Theor. Phys.}\ }\textbf {\bibinfo {volume}
  {34}},\ \bibinfo {pages} {610} (\bibinfo {year} {1971})}\BibitemShut
  {NoStop}%
\bibitem [{\citenamefont {Kosterlitz}\ and\ \citenamefont
  {Thouless}(1973)}]{Kosterlitz1973}%
  \BibitemOpen
  \bibfield  {author} {\bibinfo {author} {\bibfnamefont {D.~J.}\ \bibnamefont
  {Kosterlitz}}\ and\ \bibinfo {author} {\bibfnamefont {J.~M.}\ \bibnamefont
  {Thouless}},\ }\href {\doibase 10.1088/0022-3719/6/7/010} {\bibfield
  {journal} {\bibinfo  {journal} {J. Phys. C Solid State Phys.}\ }\textbf
  {\bibinfo {volume} {1181}},\ \bibinfo {pages} {1181} (\bibinfo {year}
  {1973})}\BibitemShut {NoStop}%
\bibitem [{\citenamefont {Hebard}\ and\ \citenamefont
  {Fiory}(1983)}]{Hebard1983}%
  \BibitemOpen
  \bibfield  {author} {\bibinfo {author} {\bibfnamefont {A.~F.}\ \bibnamefont
  {Hebard}}\ and\ \bibinfo {author} {\bibfnamefont {A.~T.}\ \bibnamefont
  {Fiory}},\ }\href {\doibase 10.1103/PhysRevLett.50.1603} {\bibfield
  {journal} {\bibinfo  {journal} {Phys. Rev. Lett.}\ }\textbf {\bibinfo
  {volume} {50}},\ \bibinfo {pages} {1603} (\bibinfo {year}
  {1983})}\BibitemShut {NoStop}%
\bibitem [{\citenamefont {Garland}\ and\ \citenamefont
  {Lee}(1987)}]{Garland1987}%
  \BibitemOpen
  \bibfield  {author} {\bibinfo {author} {\bibfnamefont {J.~C.}\ \bibnamefont
  {Garland}}\ and\ \bibinfo {author} {\bibfnamefont {H.~J.}\ \bibnamefont
  {Lee}},\ }\href {\doibase 10.1103/PhysRevB.36.3638} {\bibfield  {journal}
  {\bibinfo  {journal} {Phys. Rev. B}\ }\textbf {\bibinfo {volume} {36}},\
  \bibinfo {pages} {3638} (\bibinfo {year} {1987})}\BibitemShut {NoStop}%
\bibitem [{\citenamefont {Zhang}\ \emph {et~al.}(2014)\citenamefont {Zhang},
  \citenamefont {Sun}, \citenamefont {Zhang}, \citenamefont {Li}, \citenamefont
  {Guo}, \citenamefont {Zhao}, \citenamefont {Zhang}, \citenamefont {Peng},
  \citenamefont {Xing}, \citenamefont {Wang}, \citenamefont {Fujita},
  \citenamefont {Hirata}, \citenamefont {Li}, \citenamefont {Ding},
  \citenamefont {Tang}, \citenamefont {Wang}, \citenamefont {Wang},
  \citenamefont {He}, \citenamefont {Ji}, \citenamefont {Chen}, \citenamefont
  {Wang}, \citenamefont {Xia}, \citenamefont {Li}, \citenamefont {Wang},
  \citenamefont {Wang}, \citenamefont {Wang}, \citenamefont {Chen},
  \citenamefont {Xue},\ and\ \citenamefont {Ma}}]{Ying2014}%
  \BibitemOpen
  \bibfield  {author} {\bibinfo {author} {\bibfnamefont {W.~H.}\ \bibnamefont
  {Zhang}}, \bibinfo {author} {\bibfnamefont {Y.}~\bibnamefont {Sun}}, \bibinfo
  {author} {\bibfnamefont {J.~S.}\ \bibnamefont {Zhang}}, \bibinfo {author}
  {\bibfnamefont {F.~S.}\ \bibnamefont {Li}}, \bibinfo {author} {\bibfnamefont
  {M.~H.}\ \bibnamefont {Guo}}, \bibinfo {author} {\bibfnamefont {Y.~F.}\
  \bibnamefont {Zhao}}, \bibinfo {author} {\bibfnamefont {H.~M.}\ \bibnamefont
  {Zhang}}, \bibinfo {author} {\bibfnamefont {J.~P.}\ \bibnamefont {Peng}},
  \bibinfo {author} {\bibfnamefont {Y.}~\bibnamefont {Xing}}, \bibinfo {author}
  {\bibfnamefont {H.~C.}\ \bibnamefont {Wang}}, \bibinfo {author}
  {\bibfnamefont {T.}~\bibnamefont {Fujita}}, \bibinfo {author} {\bibfnamefont
  {A.}~\bibnamefont {Hirata}}, \bibinfo {author} {\bibfnamefont
  {Z.}~\bibnamefont {Li}}, \bibinfo {author} {\bibfnamefont {H.}~\bibnamefont
  {Ding}}, \bibinfo {author} {\bibfnamefont {C.~J.}\ \bibnamefont {Tang}},
  \bibinfo {author} {\bibfnamefont {M.}~\bibnamefont {Wang}}, \bibinfo {author}
  {\bibfnamefont {Q.~Y.}\ \bibnamefont {Wang}}, \bibinfo {author}
  {\bibfnamefont {K.}~\bibnamefont {He}}, \bibinfo {author} {\bibfnamefont
  {S.~H.}\ \bibnamefont {Ji}}, \bibinfo {author} {\bibfnamefont
  {X.}~\bibnamefont {Chen}}, \bibinfo {author} {\bibfnamefont {J.~F.}\
  \bibnamefont {Wang}}, \bibinfo {author} {\bibfnamefont {Z.~C.}\ \bibnamefont
  {Xia}}, \bibinfo {author} {\bibfnamefont {L.}~\bibnamefont {Li}}, \bibinfo
  {author} {\bibfnamefont {Y.~Y.}\ \bibnamefont {Wang}}, \bibinfo {author}
  {\bibfnamefont {J.}~\bibnamefont {Wang}}, \bibinfo {author} {\bibfnamefont
  {L.~L.}\ \bibnamefont {Wang}}, \bibinfo {author} {\bibfnamefont {M.~W.}\
  \bibnamefont {Chen}}, \bibinfo {author} {\bibfnamefont {Q.~K.}\ \bibnamefont
  {Xue}}, \ and\ \bibinfo {author} {\bibfnamefont {X.~C.}\ \bibnamefont {Ma}},\
  }\href {\doibase 10.1088/0256-307X/31/1/017401} {\bibfield  {journal}
  {\bibinfo  {journal} {Chinese Phys. Lett.}\ }\textbf {\bibinfo {volume} {31}}
  (\bibinfo {year} {2014}),\ 10.1088/0256-307X/31/1/017401}\BibitemShut
  {NoStop}%
\bibitem [{\citenamefont {Schneider}\ \emph {et~al.}(2014)\citenamefont
  {Schneider}, \citenamefont {Zaitsev}, \citenamefont {Fuchs},\ and\
  \citenamefont {{Von L{\"{o}}hneysen}}}]{Schneider2014}%
  \BibitemOpen
  \bibfield  {author} {\bibinfo {author} {\bibfnamefont {R.}~\bibnamefont
  {Schneider}}, \bibinfo {author} {\bibfnamefont {A.~G.}\ \bibnamefont
  {Zaitsev}}, \bibinfo {author} {\bibfnamefont {D.}~\bibnamefont {Fuchs}}, \
  and\ \bibinfo {author} {\bibfnamefont {H.}~\bibnamefont {{Von
  L{\"{o}}hneysen}}},\ }\href {\doibase 10.1088/0953-8984/26/45/455701}
  {\bibfield  {journal} {\bibinfo  {journal} {J. Phys. Condens. Matter}\
  }\textbf {\bibinfo {volume} {26}},\ \bibinfo {pages} {7} (\bibinfo {year}
  {2014})}\BibitemShut {NoStop}%
\bibitem [{\citenamefont {Venditti}\ \emph {et~al.}(2019)\citenamefont
  {Venditti}, \citenamefont {Biscaras}, \citenamefont {Hurand}, \citenamefont
  {Bergeal}, \citenamefont {Lesueur}, \citenamefont {Dogra}, \citenamefont
  {Budhani}, \citenamefont {Mondal}, \citenamefont {Raychaudhuri},
  \citenamefont {Caprara},\ and\ \citenamefont {Benfatto}}]{Venditti2019}%
  \BibitemOpen
  \bibfield  {author} {\bibinfo {author} {\bibfnamefont {G.}~\bibnamefont
  {Venditti}}, \bibinfo {author} {\bibfnamefont {J.}~\bibnamefont {Biscaras}},
  \bibinfo {author} {\bibfnamefont {S.}~\bibnamefont {Hurand}}, \bibinfo
  {author} {\bibfnamefont {N.}~\bibnamefont {Bergeal}}, \bibinfo {author}
  {\bibfnamefont {J.}~\bibnamefont {Lesueur}}, \bibinfo {author} {\bibfnamefont
  {A.}~\bibnamefont {Dogra}}, \bibinfo {author} {\bibfnamefont {R.~C.}\
  \bibnamefont {Budhani}}, \bibinfo {author} {\bibfnamefont {J.}~\bibnamefont
  {Mondal}, \bibfnamefont {Mintu~Jesudasan}}, \bibinfo {author} {\bibfnamefont
  {P.}~\bibnamefont {Raychaudhuri}}, \bibinfo {author} {\bibfnamefont
  {S.}~\bibnamefont {Caprara}}, \ and\ \bibinfo {author} {\bibfnamefont
  {L.}~\bibnamefont {Benfatto}},\ }\href {https://arxiv.org/abs/1905.01221}
  {\bibfield  {journal} {\bibinfo  {journal} {arXiv:1905.01221}\ } (\bibinfo
  {year} {2019})}\BibitemShut {NoStop}%
\bibitem [{\citenamefont {Bennemann}\ and\ \citenamefont
  {Ketterson}(2008)}]{Bennemann2008}%
  \BibitemOpen
  \bibfield  {author} {\bibinfo {author} {\bibfnamefont {K.~H.}\ \bibnamefont
  {Bennemann}}\ and\ \bibinfo {author} {\bibfnamefont {J.~B. J.~B.}\
  \bibnamefont {Ketterson}},\ }\href
  {https://books.google.co.uk/books?hl=en{\&}lr={\&}id=PguAgEQTiQwC{\&}oi=fnd{\&}pg=PR14{\&}dq=j+ketterson+superconductivity{\&}ots=Rcw-T9VHXm{\&}sig=4YBFPBaARxXZsvPUGbF7jeikiKA{\#}v=onepage{\&}q=j
  ketterson superconductivity{\&}f=false} {\emph {\bibinfo {title}
  {{Superconductivity}}}}\ (\bibinfo  {publisher} {Springer},\ \bibinfo {year}
  {2008})\BibitemShut {NoStop}%
\bibitem [{\citenamefont {Audouard}\ \emph {et~al.}(2015)\citenamefont
  {Audouard}, \citenamefont {Duc}, \citenamefont {Drigo}, \citenamefont
  {Toulemonde}, \citenamefont {Karlsson}, \citenamefont {Strobel},\ and\
  \citenamefont {Sulpice}}]{Audouard2015}%
  \BibitemOpen
  \bibfield  {author} {\bibinfo {author} {\bibfnamefont {A.}~\bibnamefont
  {Audouard}}, \bibinfo {author} {\bibfnamefont {F.}~\bibnamefont {Duc}},
  \bibinfo {author} {\bibfnamefont {L.}~\bibnamefont {Drigo}}, \bibinfo
  {author} {\bibfnamefont {P.}~\bibnamefont {Toulemonde}}, \bibinfo {author}
  {\bibfnamefont {S.}~\bibnamefont {Karlsson}}, \bibinfo {author}
  {\bibfnamefont {P.}~\bibnamefont {Strobel}}, \ and\ \bibinfo {author}
  {\bibfnamefont {A.}~\bibnamefont {Sulpice}},\ }\href {\doibase
  10.1209/0295-5075/109/27003} {\bibfield  {journal} {\bibinfo  {journal}
  {EPL}\ }\textbf {\bibinfo {volume} {109}} (\bibinfo {year} {2015}),\
  10.1209/0295-5075/109/27003}\BibitemShut {NoStop}%
\bibitem [{\citenamefont {Borg}\ \emph {et~al.}(2016)\citenamefont {Borg},
  \citenamefont {Zhou}, \citenamefont {Eckberg}, \citenamefont {Campbell},
  \citenamefont {Saha}, \citenamefont {Paglione},\ and\ \citenamefont
  {Rodriguez}}]{Borg2016a}%
  \BibitemOpen
  \bibfield  {author} {\bibinfo {author} {\bibfnamefont {C.~K.}\ \bibnamefont
  {Borg}}, \bibinfo {author} {\bibfnamefont {X.}~\bibnamefont {Zhou}}, \bibinfo
  {author} {\bibfnamefont {C.}~\bibnamefont {Eckberg}}, \bibinfo {author}
  {\bibfnamefont {D.~J.}\ \bibnamefont {Campbell}}, \bibinfo {author}
  {\bibfnamefont {S.~R.}\ \bibnamefont {Saha}}, \bibinfo {author}
  {\bibfnamefont {J.}~\bibnamefont {Paglione}}, \ and\ \bibinfo {author}
  {\bibfnamefont {E.~E.}\ \bibnamefont {Rodriguez}},\ }\href {\doibase
  10.1103/PhysRevB.93.094522} {\bibfield  {journal} {\bibinfo  {journal} {Phys.
  Rev. B}\ }\textbf {\bibinfo {volume} {93}},\ \bibinfo {pages} {94522}
  (\bibinfo {year} {2016})}\BibitemShut {NoStop}%
\bibitem [{\citenamefont {Werthamer}\ \emph {et~al.}(1966)\citenamefont
  {Werthamer}, \citenamefont {Helfand},\ and\ \citenamefont
  {Hohenberg}}]{Werthamer1966}%
  \BibitemOpen
  \bibfield  {author} {\bibinfo {author} {\bibfnamefont {N.~R.}\ \bibnamefont
  {Werthamer}}, \bibinfo {author} {\bibfnamefont {E.}~\bibnamefont {Helfand}},
  \ and\ \bibinfo {author} {\bibfnamefont {P.~C.}\ \bibnamefont {Hohenberg}},\
  }\href {\doibase 10.1103/PhysRev.147.295} {\bibfield  {journal} {\bibinfo
  {journal} {Phys. Rev.}\ }\textbf {\bibinfo {volume} {147}},\ \bibinfo {pages}
  {295} (\bibinfo {year} {1966})}\BibitemShut {NoStop}%
\bibitem [{\citenamefont {Khim}\ \emph {et~al.}(2010)\citenamefont {Khim},
  \citenamefont {Kim}, \citenamefont {Choi}, \citenamefont {Bang},
  \citenamefont {Nohara}, \citenamefont {Takagi},\ and\ \citenamefont
  {Kim}}]{Khim2010}%
  \BibitemOpen
  \bibfield  {author} {\bibinfo {author} {\bibfnamefont {S.}~\bibnamefont
  {Khim}}, \bibinfo {author} {\bibfnamefont {J.~W.}\ \bibnamefont {Kim}},
  \bibinfo {author} {\bibfnamefont {E.~S.}\ \bibnamefont {Choi}}, \bibinfo
  {author} {\bibfnamefont {Y.}~\bibnamefont {Bang}}, \bibinfo {author}
  {\bibfnamefont {M.}~\bibnamefont {Nohara}}, \bibinfo {author} {\bibfnamefont
  {H.}~\bibnamefont {Takagi}}, \ and\ \bibinfo {author} {\bibfnamefont {K.~H.}\
  \bibnamefont {Kim}},\ }\href {\doibase 10.1103/PhysRevB.81.184511} {\bibfield
   {journal} {\bibinfo  {journal} {Phys. Rev. B}\ }\textbf {\bibinfo {volume}
  {81}},\ \bibinfo {pages} {184511} (\bibinfo {year} {2010})}\BibitemShut
  {NoStop}%
\bibitem [{\citenamefont {Tinkham}(1996)}]{Tinkham1996}%
  \BibitemOpen
  \bibfield  {author} {\bibinfo {author} {\bibfnamefont {M.}~\bibnamefont
  {Tinkham}},\ }in\ \href {\doibase 10.1201/9780203738887-1} {\emph {\bibinfo
  {booktitle} {Introduction to Superconductivity}}}\ (\bibinfo  {publisher}
  {Dover Publications},\ \bibinfo {year} {1996})\BibitemShut {NoStop}%
\bibitem [{\citenamefont {Clogston}(1962)}]{Clogston1962}%
  \BibitemOpen
  \bibfield  {author} {\bibinfo {author} {\bibfnamefont {A.~M.}\ \bibnamefont
  {Clogston}},\ }\href {\doibase 10.1103/PhysRevLett.9.266} {\bibfield
  {journal} {\bibinfo  {journal} {Phys. Rev. Lett.}\ }\textbf {\bibinfo
  {volume} {9}},\ \bibinfo {pages} {266} (\bibinfo {year} {1962})}\BibitemShut
  {NoStop}%
\bibitem [{\citenamefont {Lee}\ \emph {et~al.}(1997)\citenamefont {Lee},
  \citenamefont {Naughton}, \citenamefont {Danner},\ and\ \citenamefont
  {Chaikin}}]{Lee1997}%
  \BibitemOpen
  \bibfield  {author} {\bibinfo {author} {\bibfnamefont {I.~J.}\ \bibnamefont
  {Lee}}, \bibinfo {author} {\bibfnamefont {M.~J.}\ \bibnamefont {Naughton}},
  \bibinfo {author} {\bibfnamefont {G.~M.}\ \bibnamefont {Danner}}, \ and\
  \bibinfo {author} {\bibfnamefont {P.~M.}\ \bibnamefont {Chaikin}},\ }\href
  {\doibase 10.1103/PhysRevLett.78.3555} {\bibfield  {journal} {\bibinfo
  {journal} {Phys. Rev. Lett.}\ }\textbf {\bibinfo {volume} {78}},\ \bibinfo
  {pages} {3555} (\bibinfo {year} {1997})}\BibitemShut {NoStop}%
\bibitem [{\citenamefont {Fulde}\ and\ \citenamefont
  {Ferrell}(1964)}]{Fulde1964}%
  \BibitemOpen
  \bibfield  {author} {\bibinfo {author} {\bibfnamefont {P.}~\bibnamefont
  {Fulde}}\ and\ \bibinfo {author} {\bibfnamefont {R.~A.}\ \bibnamefont
  {Ferrell}},\ }\href {\doibase 10.1103/PhysRev.135.A550} {\bibfield  {journal}
  {\bibinfo  {journal} {Phys. Rev.}\ }\textbf {\bibinfo {volume} {135}},\
  \bibinfo {pages} {A550} (\bibinfo {year} {1964})}\BibitemShut {NoStop}%
\bibitem [{\citenamefont {Larkin}\ and\ \citenamefont
  {Ovchinnikov}(1964)}]{Larkin1964}%
  \BibitemOpen
  \bibfield  {author} {\bibinfo {author} {\bibfnamefont {A.~I.}\ \bibnamefont
  {Larkin}}\ and\ \bibinfo {author} {\bibfnamefont {Y.~N.}\ \bibnamefont
  {Ovchinnikov}},\ }\href@noop {} {\bibfield  {journal} {\bibinfo  {journal}
  {Zh. Eksp. Teor. Fiz.}\ }\textbf {\bibinfo {volume} {47}},\ \bibinfo {pages}
  {1136} (\bibinfo {year} {1964})}\BibitemShut {NoStop}%
\bibitem [{\citenamefont {{J. M. Lu}}\ \emph {et~al.}(2015)\citenamefont {{J.
  M. Lu}}, \citenamefont {{O. Zheliuk}}, \citenamefont {{I. Leermakers}},
  \citenamefont {{N. F. Q. Yuan}}, \citenamefont {{U. Zeitler}}, \citenamefont
  {{K. T. Law}},\ and\ \citenamefont {{J. T. Ye}}}]{Lu2015}%
  \BibitemOpen
  \bibfield  {author} {\bibinfo {author} {\bibnamefont {{J. M. Lu}}}, \bibinfo
  {author} {\bibnamefont {{O. Zheliuk}}}, \bibinfo {author} {\bibnamefont {{I.
  Leermakers}}}, \bibinfo {author} {\bibnamefont {{N. F. Q. Yuan}}}, \bibinfo
  {author} {\bibnamefont {{U. Zeitler}}}, \bibinfo {author} {\bibnamefont {{K.
  T. Law}}}, \ and\ \bibinfo {author} {\bibnamefont {{J. T. Ye}}},\ }\href
  {\doibase 10.1126/science.aab2277} {\bibfield  {journal} {\bibinfo  {journal}
  {Science}\ }\textbf {\bibinfo {volume} {350}},\ \bibinfo {pages} {1353}
  (\bibinfo {year} {2015})}\BibitemShut {NoStop}%
\bibitem [{\citenamefont {Sprau}\ \emph {et~al.}(2017)\citenamefont {Sprau},
  \citenamefont {Kostin}, \citenamefont {Kreisel}, \citenamefont {B{\"o}hmer},
  \citenamefont {Taufour}, \citenamefont {Canfield}, \citenamefont {Mukherjee},
  \citenamefont {Hirschfeld}, \citenamefont {Andersen},\ and\ \citenamefont
  {Davis}}]{Sprau2017}%
  \BibitemOpen
  \bibfield  {author} {\bibinfo {author} {\bibfnamefont {P.~O.}\ \bibnamefont
  {Sprau}}, \bibinfo {author} {\bibfnamefont {A.}~\bibnamefont {Kostin}},
  \bibinfo {author} {\bibfnamefont {A.}~\bibnamefont {Kreisel}}, \bibinfo
  {author} {\bibfnamefont {A.~E.}\ \bibnamefont {B{\"o}hmer}}, \bibinfo
  {author} {\bibfnamefont {V.}~\bibnamefont {Taufour}}, \bibinfo {author}
  {\bibfnamefont {P.~C.}\ \bibnamefont {Canfield}}, \bibinfo {author}
  {\bibfnamefont {S.}~\bibnamefont {Mukherjee}}, \bibinfo {author}
  {\bibfnamefont {P.~J.}\ \bibnamefont {Hirschfeld}}, \bibinfo {author}
  {\bibfnamefont {B.~M.}\ \bibnamefont {Andersen}}, \ and\ \bibinfo {author}
  {\bibfnamefont {J.~C.~S.}\ \bibnamefont {Davis}},\ }\href {\doibase
  10.1126/science.aal1575} {\bibfield  {journal} {\bibinfo  {journal}
  {Science}\ }\textbf {\bibinfo {volume} {357}},\ \bibinfo {pages} {75}
  (\bibinfo {year} {2017})},\ \Eprint
  {http://arxiv.org/abs/https://science.sciencemag.org/content/357/6346/75.full.pdf}
  {https://science.sciencemag.org/content/357/6346/75.full.pdf} \BibitemShut
  {NoStop}%
\bibitem [{\citenamefont {Gurevich}(2010)}]{Gurevich2010}%
  \BibitemOpen
  \bibfield  {author} {\bibinfo {author} {\bibfnamefont {A.}~\bibnamefont
  {Gurevich}},\ }\href {\doibase 10.1103/PhysRevB.82.184504} {\bibfield
  {journal} {\bibinfo  {journal} {Phys. Rev. B}\ }\textbf {\bibinfo {volume}
  {82}},\ \bibinfo {pages} {184504} (\bibinfo {year} {2010})}\BibitemShut
  {NoStop}%
\bibitem [{\citenamefont {Gurevich}(2014)}]{Gurevich2014}%
  \BibitemOpen
  \bibfield  {author} {\bibinfo {author} {\bibfnamefont {A.}~\bibnamefont
  {Gurevich}},\ }\href {\doibase 10.1146/annurev-conmatphys-031113-133822}
  {\bibfield  {journal} {\bibinfo  {journal} {Annual Review of Condensed Matter
  Physics}\ }\textbf {\bibinfo {volume} {5}},\ \bibinfo {pages} {35} (\bibinfo
  {year} {2014})},\ \Eprint
  {http://arxiv.org/abs/https://doi.org/10.1146/annurev-conmatphys-031113-133822}
  {https://doi.org/10.1146/annurev-conmatphys-031113-133822} \BibitemShut
  {NoStop}%
\bibitem [{\citenamefont {Terashima}\ \emph {et~al.}(2014)\citenamefont
  {Terashima}, \citenamefont {Kikugawa}, \citenamefont {Kiswandhi},
  \citenamefont {Choi}, \citenamefont {Brooks}, \citenamefont {Kasahara},
  \citenamefont {Watashige}, \citenamefont {Ikeda}, \citenamefont {Shibauchi},
  \citenamefont {Matsuda}, \citenamefont {Wolf}, \citenamefont {B\"ohmer},
  \citenamefont {Hardy}, \citenamefont {Meingast}, \citenamefont {L\"ohneysen},
  \citenamefont {Suzuki}, \citenamefont {Arita},\ and\ \citenamefont
  {Uji}}]{Terashima2014}%
  \BibitemOpen
  \bibfield  {author} {\bibinfo {author} {\bibfnamefont {T.}~\bibnamefont
  {Terashima}}, \bibinfo {author} {\bibfnamefont {N.}~\bibnamefont {Kikugawa}},
  \bibinfo {author} {\bibfnamefont {A.}~\bibnamefont {Kiswandhi}}, \bibinfo
  {author} {\bibfnamefont {E.-S.}\ \bibnamefont {Choi}}, \bibinfo {author}
  {\bibfnamefont {J.~S.}\ \bibnamefont {Brooks}}, \bibinfo {author}
  {\bibfnamefont {S.}~\bibnamefont {Kasahara}}, \bibinfo {author}
  {\bibfnamefont {T.}~\bibnamefont {Watashige}}, \bibinfo {author}
  {\bibfnamefont {H.}~\bibnamefont {Ikeda}}, \bibinfo {author} {\bibfnamefont
  {T.}~\bibnamefont {Shibauchi}}, \bibinfo {author} {\bibfnamefont
  {Y.}~\bibnamefont {Matsuda}}, \bibinfo {author} {\bibfnamefont
  {T.}~\bibnamefont {Wolf}}, \bibinfo {author} {\bibfnamefont {A.~E.}\
  \bibnamefont {B\"ohmer}}, \bibinfo {author} {\bibfnamefont {F.}~\bibnamefont
  {Hardy}}, \bibinfo {author} {\bibfnamefont {C.}~\bibnamefont {Meingast}},
  \bibinfo {author} {\bibfnamefont {H.~v.}\ \bibnamefont {L\"ohneysen}},
  \bibinfo {author} {\bibfnamefont {M.-T.}\ \bibnamefont {Suzuki}}, \bibinfo
  {author} {\bibfnamefont {R.}~\bibnamefont {Arita}}, \ and\ \bibinfo {author}
  {\bibfnamefont {S.}~\bibnamefont {Uji}},\ }\href {\doibase
  10.1103/PhysRevB.90.144517} {\bibfield  {journal} {\bibinfo  {journal} {Phys.
  Rev. B}\ }\textbf {\bibinfo {volume} {90}},\ \bibinfo {pages} {144517}
  (\bibinfo {year} {2014})}\BibitemShut {NoStop}%
\bibitem [{\citenamefont {Anderson}(1958)}]{Anderson1958}%
  \BibitemOpen
  \bibfield  {author} {\bibinfo {author} {\bibfnamefont {P.~W.}\ \bibnamefont
  {Anderson}},\ }\href {\doibase 10.1103/PhysRev.109.1492} {\bibfield
  {journal} {\bibinfo  {journal} {Phys. Rev.}\ }\textbf {\bibinfo {volume}
  {109}},\ \bibinfo {pages} {1492} (\bibinfo {year} {1958})}\BibitemShut
  {NoStop}%
\bibitem [{\citenamefont {{Markovic, N, C. Christiansen A. M. Mack, W. H.
  Huber}}(1999)}]{Gold1999}%
  \BibitemOpen
  \bibfield  {author} {\bibinfo {author} {\bibfnamefont {A.~M.~G.}\
  \bibnamefont {{Markovic, N, C. Christiansen A. M. Mack, W. H. Huber}}},\
  }\href {\doibase 10.1002/pssb.200983017} {\bibfield  {journal} {\bibinfo
  {journal} {Phys. Status Solidi Basic Res.}\ }\textbf {\bibinfo {volume}
  {60}},\ \bibinfo {pages} {4320} (\bibinfo {year} {1999})}\BibitemShut
  {NoStop}%
\bibitem [{\citenamefont {Shiogai}\ \emph {et~al.}(2018)\citenamefont
  {Shiogai}, \citenamefont {Kimura}, \citenamefont {Awaji}, \citenamefont
  {Nojima},\ and\ \citenamefont {Tsukazaki}}]{Shiogai2018}%
  \BibitemOpen
  \bibfield  {author} {\bibinfo {author} {\bibfnamefont {J.}~\bibnamefont
  {Shiogai}}, \bibinfo {author} {\bibfnamefont {S.}~\bibnamefont {Kimura}},
  \bibinfo {author} {\bibfnamefont {S.}~\bibnamefont {Awaji}}, \bibinfo
  {author} {\bibfnamefont {T.}~\bibnamefont {Nojima}}, \ and\ \bibinfo {author}
  {\bibfnamefont {A.}~\bibnamefont {Tsukazaki}},\ }\href {\doibase
  10.1103/PhysRevB.97.174520} {\bibfield  {journal} {\bibinfo  {journal} {Phys.
  Rev. B}\ }\textbf {\bibinfo {volume} {97}},\ \bibinfo {pages} {174520}
  (\bibinfo {year} {2018})}\BibitemShut {NoStop}%
\bibitem [{\citenamefont {Blatter}\ \emph {et~al.}(1994)\citenamefont
  {Blatter}, \citenamefont {Feigel'man}, \citenamefont {Geshkenbein},
  \citenamefont {Larkin},\ and\ \citenamefont {Vinokur}}]{Blatter1994}%
  \BibitemOpen
  \bibfield  {author} {\bibinfo {author} {\bibfnamefont {G.}~\bibnamefont
  {Blatter}}, \bibinfo {author} {\bibfnamefont {M.~V.}\ \bibnamefont
  {Feigel'man}}, \bibinfo {author} {\bibfnamefont {V.~B.}\ \bibnamefont
  {Geshkenbein}}, \bibinfo {author} {\bibfnamefont {A.~I.}\ \bibnamefont
  {Larkin}}, \ and\ \bibinfo {author} {\bibfnamefont {V.~M.}\ \bibnamefont
  {Vinokur}},\ }\href {\doibase 10.1103/RevModPhys.66.1125} {\bibfield
  {journal} {\bibinfo  {journal} {Rev. Mod. Phys.}\ }\textbf {\bibinfo {volume}
  {66}},\ \bibinfo {pages} {1125} (\bibinfo {year} {1994})}\BibitemShut
  {NoStop}%
\bibitem [{\citenamefont {Huang}\ and\ \citenamefont
  {Hoffman}(2017)}]{Huang2017}%
  \BibitemOpen
  \bibfield  {author} {\bibinfo {author} {\bibfnamefont {D.}~\bibnamefont
  {Huang}}\ and\ \bibinfo {author} {\bibfnamefont {J.~E.}\ \bibnamefont
  {Hoffman}},\ }\href {\doibase 10.1146/annurev-conmatphys-031016-025242}
  {\bibfield  {journal} {\bibinfo  {journal} {Annual Review of Condensed Matter
  Physics}\ }\textbf {\bibinfo {volume} {8}},\ \bibinfo {pages} {311} (\bibinfo
  {year} {2017})},\ \Eprint
  {http://arxiv.org/abs/https://doi.org/10.1146/annurev-conmatphys-031016-025242}
  {https://doi.org/10.1146/annurev-conmatphys-031016-025242} \BibitemShut
  {NoStop}%
\bibitem [{\citenamefont {Tinkham}(1963)}]{Tinkham1963}%
  \BibitemOpen
  \bibfield  {author} {\bibinfo {author} {\bibfnamefont {M.}~\bibnamefont
  {Tinkham}},\ }\href {\doibase 10.1103/PhysRev.129.2413} {\bibfield  {journal}
  {\bibinfo  {journal} {Phys. Rev.}\ }\textbf {\bibinfo {volume} {129}},\
  \bibinfo {pages} {2413} (\bibinfo {year} {1963})}\BibitemShut {NoStop}%
\end{thebibliography}%


\newpage
\clearpage

\newcommand{\blue}{\textcolor{blue}}
\newcommand{\bdm}[1]{\mbox{\boldmath $#1$}}

\renewcommand{\thefigure}{S\arabic{figure}} 
\renewcommand{\thetable}{S\arabic{table}} 

\newlength{\figwidth}
\figwidth=0.48\textwidth

\setcounter{figure}{0}

\newcommand{\fig}[3]
{
\begin{figure}[!tb]
\vspace*{-0.1cm}
\[
\includegraphics[width=\figwidth]{#1}
\]
\vskip -0.2cm
\caption{\label{#2}
\small#3
}
\end{figure}}

\newpage
\clearpage

{\bf Upper critical field and relevant parameters}

In an conventional BCS superconductor, an external magnetic field can destroy the Cooper pairs
either due to a)  the orbital pair breaking due to the Lorentz force acting on the charge of the paired electrons,
known as the orbital limit, $\Delta \sim \mu_0 \mu_B H_{\rm c}^{orb}$
or b)  due to the Pauli paramagnetic pair breaking
as a result of the Zeeman energy that leads to the alignment
of the two opposite spins of the two electrons forming the singlet state with the applied field,
 called the Pauli paramagnetic limit \cite{Tinkham1963}.

{\bf WHH model}

The single-band model by WHH provides predictions for the upper critical field, $H_{c2}$ of type-II superconductors as a function of temperature, $T$
in the dirty limit given by \cite{Werthamer1966}:

\begin{equation}
\ln\frac{1}{t}=\sum_{\nu=-\infty}^{\infty}\frac{1}{|{2\nu+1}|}-
\nonumber
\label{eq:WHH}
\end{equation}

\begin{equation}
{\left[|{2\nu+1}+\frac{h}{t}+\frac{(\alpha_{\rm M} ht)^2}{|{2\nu +1}+(h+\lambda_{so}/t)}\right]}^{-1},
\label{eq:WHH}
\end{equation}

\noindent where $t=T/T_{\rm c}$, and
$h=4H_{\rm c2}/[\pi^2T_c\left(\mathrm{d}H_{\rm  c2}/{\mathrm{d}T}\right)_{T=T_{\rm c}}$.
$\lambda_{\rm so}=\hbar/(3\pi k_B T_c\tau_{so})$ accounts for spin-orbit and spin-flip scattering
with $\tau_{\rm so}$ as the mean free scattering time.

In the absence of spin paramagnetic effect ($\alpha_{\rm M}=0$), the upper critical field is
restricted by orbital pair breaking effect. In the weak-coupling case, assuming $\lambda_{so} =0$):
\begin{equation}
  H^{orb}_{\rm c2} = -0.69 \cdot T_{\rm c} \cdot |\textrm{d}H_{c2}/\textrm{d}T |_{T=T_{c}}
\end{equation}

The Maki parameter,
$\alpha_{\rm M}=\sqrt{2}H_{\rm c2}^{orb}/H_{\rm c2}^{pm}$
indicates which of the orbital or spin-paramagnetic effects is more dominant in determining the upper critical field.
When $\alpha_{\rm M}> 1$ the paramagnetic effects
become essential.
This condition
can be easily satisfied for materials with low Fermi energies and high $T_{\rm c}$.

In a single-band system, in the clean limit the
parameter $\alpha_{\rm M}$
for $H||c$ is given by \cite{Gurevich2014}:
\begin{equation}
\alpha_{\rm M} = \frac{\pi \Delta m^*}{4E_{\rm F } m_e}
\end{equation}
and for $H||$($ab$) by
\begin{equation}
\alpha_{\rm M} = \frac{\pi \Delta v^{ab}_{\rm F}m^*}{4E_{\rm F } m_e v^c_{\rm F}}
\end{equation}
where where  $v^{ab}_{\rm F}$ and $v^{c}_{\rm F}$
are Fermi surface velocities.

 For $H||c$, the
FFLO instability occurs at $\alpha_{\rm M} >1.8$, where
for a single-band spherical Fermi surface
which implies $\Delta > E_{\rm F}$, which can hardly
happen in a single-band conventional superconductor with
a low effective mass $m^* \sim m_e$ \cite{Gurevich2010}.
The criterion $\alpha_{\rm M} >1.8$ can be
satisfied more easily in strongly anisotropic materials in a magnetic field parallel to
the layers, $H||ab$, in which case $\alpha_{\rm M} $ is enhanced by the large
$v_{\rm F}^{ab}/v_{\rm F}^c$
ratio. A FFLO state would be easier
to stabilize for $H||ab$,  in strongly correlated materials
 in presence of strong correlations where $m^* \gg m_e$ \cite{Gurevich2010}.

{\bf Coherence lengths.}
The superconducting coherence length
 quantifies the size of Cooper pairs. 
In a clean metal with a large Fermi velocity, it is the strong overlap of Cooper pairs that provides
the superconducting phase coherence and the coherence length is given by:

\begin{equation}
\xi_0 = \frac{\hbar v_{\rm F}}{\pi \Delta}
\end{equation}
where $v_{\rm F}$ is the Fermi velocity
and $\Delta$ the energy gap.

In a dirty superconductor,
the mean free path $\ell$ is much shorter than the coherence length
which is given by:

\begin{equation}
\xi_d = \left (\frac{\hbar \ell v_{\rm F}}{3 \pi \Delta}\right)^{1/2}
\end{equation}
where $v_{\rm F}$ is the Fermi velocity
and $\Delta$ the energy gap.

The upper critical field perpendicular
to the conducting planes, $H_{\rm c2 \perp}$ is determined by
vortices whose screening currents flow parallel to the
planes.

The coherent lengths $\xi_{ab}$ and $\xi_{c}$ were estimated from the single band anisotropic Ginzburg-Landau equations
\begin{equation}
H_{\rm c2}^{c}=H_{\rm c2 \perp}=\frac{\Phi_0}{2 \pi \xi_{ab}^2 }
	\end{equation}

and
\begin{equation}
H_{\rm c2}^{ab}=H_{\rm c2||}=\frac{\Phi_0}{2 \pi \xi_{ab} \xi_{c}}
	\end{equation}

The anisotropy ratio $\Gamma $ is given by

\begin{equation}
\Gamma =\frac{H_{\rm c2}^{ab}}{H_{\rm c2}^{c}}=\frac{H_{\rm c2||}}{H_{\rm c2 \perp}}
=   \frac{\xi_{ab}}{ \xi_{c}}
	\end{equation}

Superconductivity in anisotropic superconductors
is confined to the two-dimensional planes.
The three-dimensional phase coherence is
provided by Josephson current between planes.

{\bf 2D Ginzburg-Landau model}
A 2D Ginzburg-Landau (2D-GL) theory predicts that the upper critical field follows the following relationship:
		\begin{equation}
	H_{\rm c2}^{\perp}(T)=\frac{\varphi_{0} \sqrt{12}}{2 \pi \xi_{G L}(0) t}\left(1-\frac{T}{T_{c}}\right)^{1 / 2}
	\label{eqn:Hc2}	
	\end{equation}
	where $\phi_{0}$ is the flux quantum, $t$ is the thickness of the superconducting layer, and $\xi_{\rm GL}$ is the Ginzburg-Landau coherence length.
The 2D-GL behaviour exhibits a square root temperature dependence of $H_{c2} \parallel_{\rm ab}$ close to $T_{\rm c}$
which accurately describes the observed behaviour of the flake with $t$= 14~nm.

\vspace{0.5cm}

{\bf Angular dependence of the upper critical field}
The  anisotropic Ginzburg-Landau (GL) theory \cite{Bennemann2008} describes
the angular dependence of the upper critical field
given by the following expression:
\begin{equation}
\left(\frac{H_{\rm c2}(\theta) \sin \theta}{H_{c2}^{\perp}}\right)^2 + \left(\frac{H_{c2}(\theta) \cos \theta}{H_{c2}^{\parallel}}\right)^2 = 1.
\label{eq:GL}
\end{equation}

When the conducting layers
consisting of a quasi-2D superconductors are sufficiently decoupled, the angle dependence of $H_{\rm c2}$
obeys the 2D Tinkham model  and the orbital pair-breaking effect is dominant.
However, when $H_{\rm c2}$ is dominated by the Pauli paramagnetic pair-breaking effect, the angle
dependence is expressed by the 3D anisotropic GL model regardless of the interlayer coupling strength.
In this case, the anisotropy of superconductivity reflects that of  the $g$-factor in the Zeeman
term.

The 2D Tinkham model for thin-film superconductivity is given by \cite{Tinkham1963},

\begin{equation}
\frac{H_{\rm c2}(\theta) | \cos \theta| }{H_{c2}^{\perp}} + \left(\frac{H_{c2}(\theta) \sin \theta}{H_{c2}^{\parallel}}\right)^2 = 1.
\label{eq:TK}
\end{equation}

\begin{table*}[htbp]
\caption{ Summary of superconducting parameters
of the FeSe thin flake devices for different thickness $t$.
 The listed parameters contain  beside superconducting critical temperature, $T_{c}$,
  the WHH fitting parameters related to the the upper critical field slope of near $T_{\rm c}$,
 (\textrm{d}\textit{H$_{\rm c2}$/}\textrm{d}\textit{T)$_{T=T_{\rm c}}$},
the orbital upper critical field, \textit{$\mu_0$H$^{orb}_{\rm c2}$},
the experimental upper critical field \textit{$\mu_{0}$H$_{\rm c2}$}.
The extracted Maki parameter, $\alpha_{\rm M}$, and the spin-orbit scattering constant, $\lambda_{SO}$, are listed.
}
\centering
\begin{tabular}{C{1cm}C{1cm}C{1cm}C{1cm}C{1cm}C{1cm}C{1cm}C{1cm}C{1cm}C{1cm}C{1cm}C{1cm}}
    \hline
     \hline
 \textit{t}
 & \textit{$T_{\rm c}$}
 & \multicolumn{2}{c}{(\textrm{d}\textit{$H_{\rm c2}$/}\textrm{d}\textit{T)$_{T=T_{\rm c}}$} }
 & \multicolumn{2}{c}{$\mu_{0} H^{\rm orb}_{\rm c2}$(0)}
 & \multicolumn{2}{c}{$\mu_{0}H_{\rm c2}$(0)}
 & \multicolumn{2}{c}{$\alpha_{\rm M}$}
 & \multicolumn{2}{c}{$\lambda_{\rm SO}$}
 \\
 (nm)
 &(K)
 & \multicolumn{2}{c}{(T/K)}
 & \multicolumn{2}{c}{(T)}
 & \multicolumn{2}{c}{(T)}
 & \multicolumn{2}{c}{}
 & \multicolumn{2}{c}{}
 \\
 \hline
  &&\textit{H$\parallel$c}&\textit{H$\parallel$ab}&\textit{H$\parallel$c}&\textit{H$\parallel$ab}&\textit{H$\parallel$c}&\textit{H$\parallel$ab}&\textit{H$\parallel$c}&\textit{H$\parallel$ab}&\textit{H$\parallel$c}&\textit{H$\parallel$ab}\\ \hline
 100&7.02&2.85&9.21&14.0&42.4&14.2&20.1&0&2.40&0.20&0.20\\
 58 &6.85&2.94&8.98&13.9&38.4&13.6&19.9&0&2.40&0.20&0.20\\
 24 &5.38&2.07&13.5&7.69&42.6&7.71&16.1&0&4.15&0.30&0.30\\
 14 &3.63&1.31&37.5&3.02&96.0&3.29&13.3&0&11.0&0.35&0.35\\
  \hline
   \hline
\end{tabular}
\end{table*}

\begin{figure*}[htbp]
	\centering
		   \includegraphics[trim={0cm 0cm 0cm 0cm}, width=0.8\linewidth,clip=true]{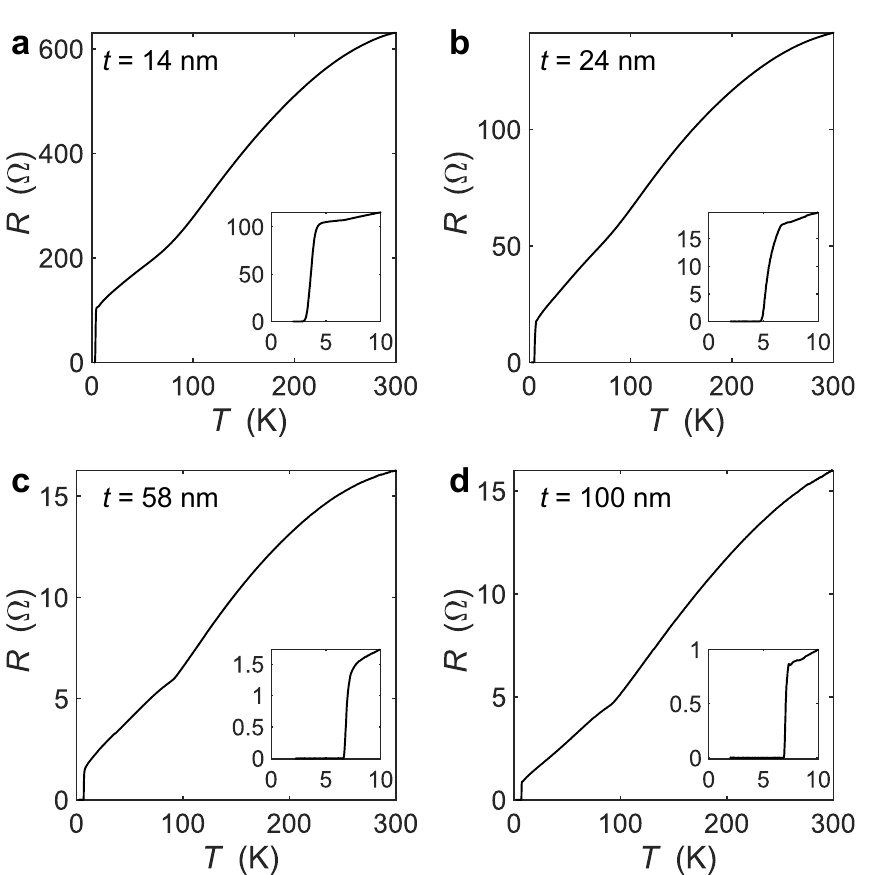}
	\caption{a-d) Temperature  dependence of
		resistance four FeSe thin flakes with different thickness used to build the phase diagram in Fig. \ref{fig:anisotropy}.
		The insets show a close up of the superconducting transitions. }
	\label{fig:RvsTSM}
\end{figure*}

\begin{figure*}[htbp]
	\centering
	\includegraphics[trim={0cm 0cm 0cm 0cm}, width=0.8\linewidth,clip=true]{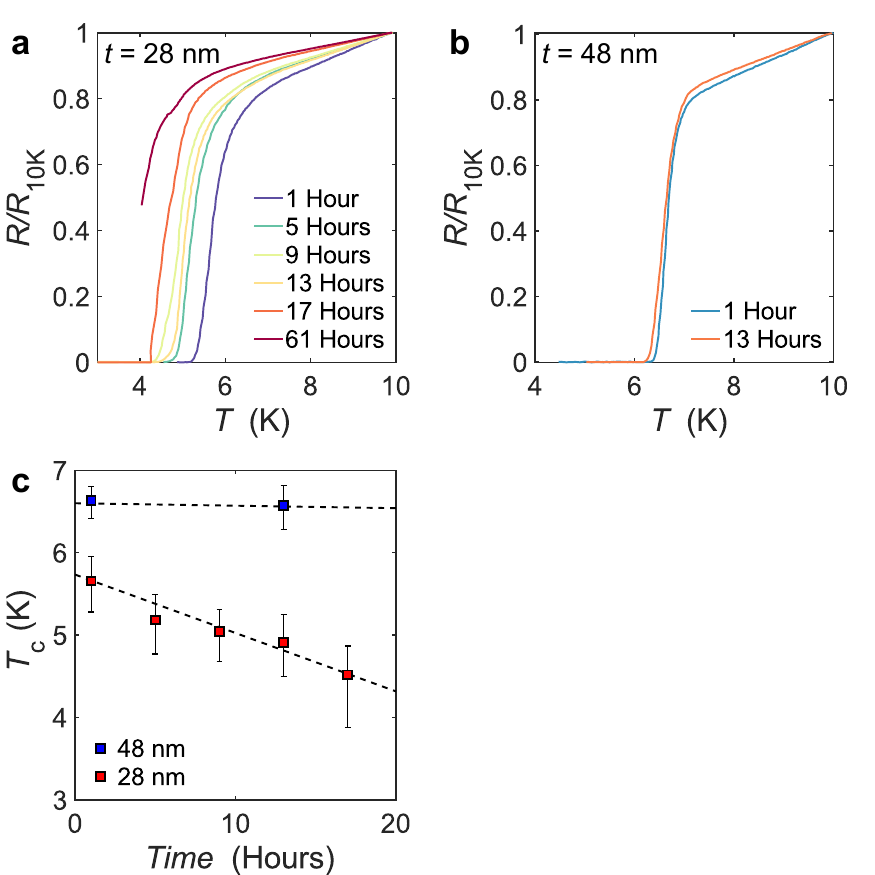}
	\caption{a,b) Temperature  dependence of
resistance normalised to 10 K ($R/R_{\rm{10K}}$) for a $t$=28~nm and $t$=48~nm FeSe device measured after exposure to air for several hours. Samples were measured over several days with no effort to control the humidity. The temperature during exposure was kept constant at 300 K. b) Extracted superconducting transition temperatures from a,b), here the error bars indicate the width of the resistive transition. The dashed lines are guides to the eye. }
	\label{fig:RvsT_and_degradation}
\end{figure*}

\begin{figure*}[htbp]
	\centering
	\includegraphics[trim={0cm 0cm 0cm 0cm}, width=0.8\linewidth,clip=true]{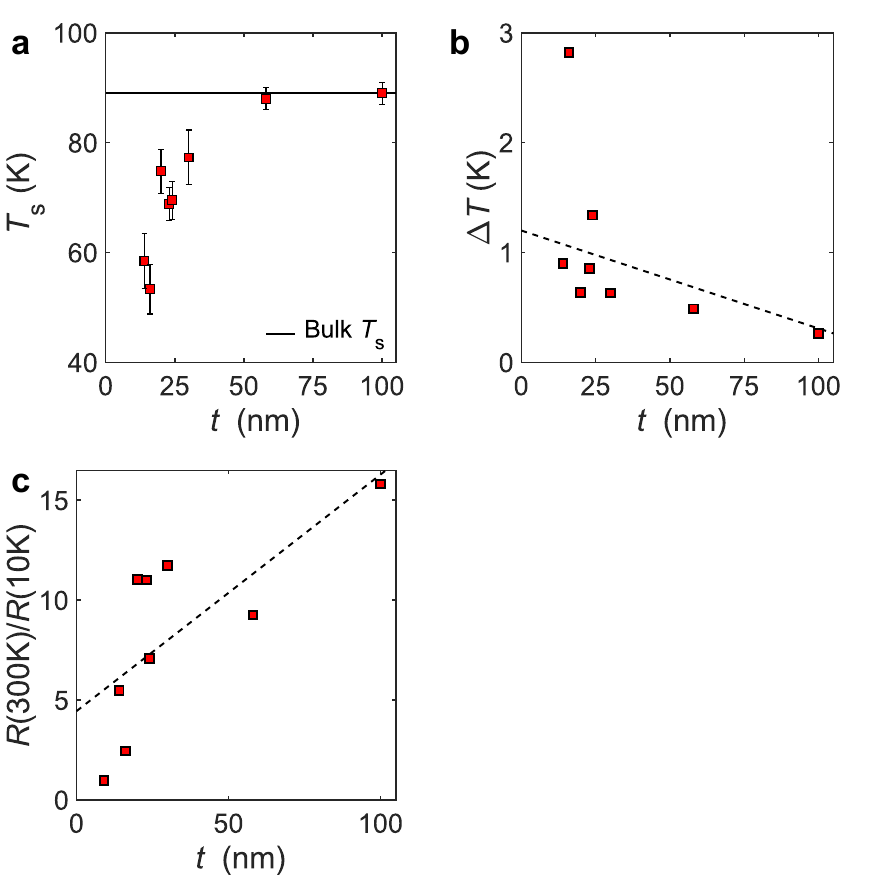}
	\caption{a) Thickness dependence of the minimum in $\rm{d}\it{R}/\rm{d}\it{T}$ associated with the temperature at which bulk FeSe undergoes nematic ordering. The solid line indicated the temperature at which this structural transition happens in bulk FeSe single crystals ($T_{\rm{s}}$=89~K).	
	b) Thickness dependent of the superconducting transition width. The dashed line is a guide to the eye.
	c) Thickness dependence of the residual resistivity ratio ($R/R_{\rm{300K}}$). The dashed line is a guide to the eye.	}
	\label{fig:RvsT_parameters}
\end{figure*}

\begin{figure*}[ht]
 \centering
  	\includegraphics[trim={0cm 0cm 0cm 0cm}, width=0.9\linewidth,clip=true]{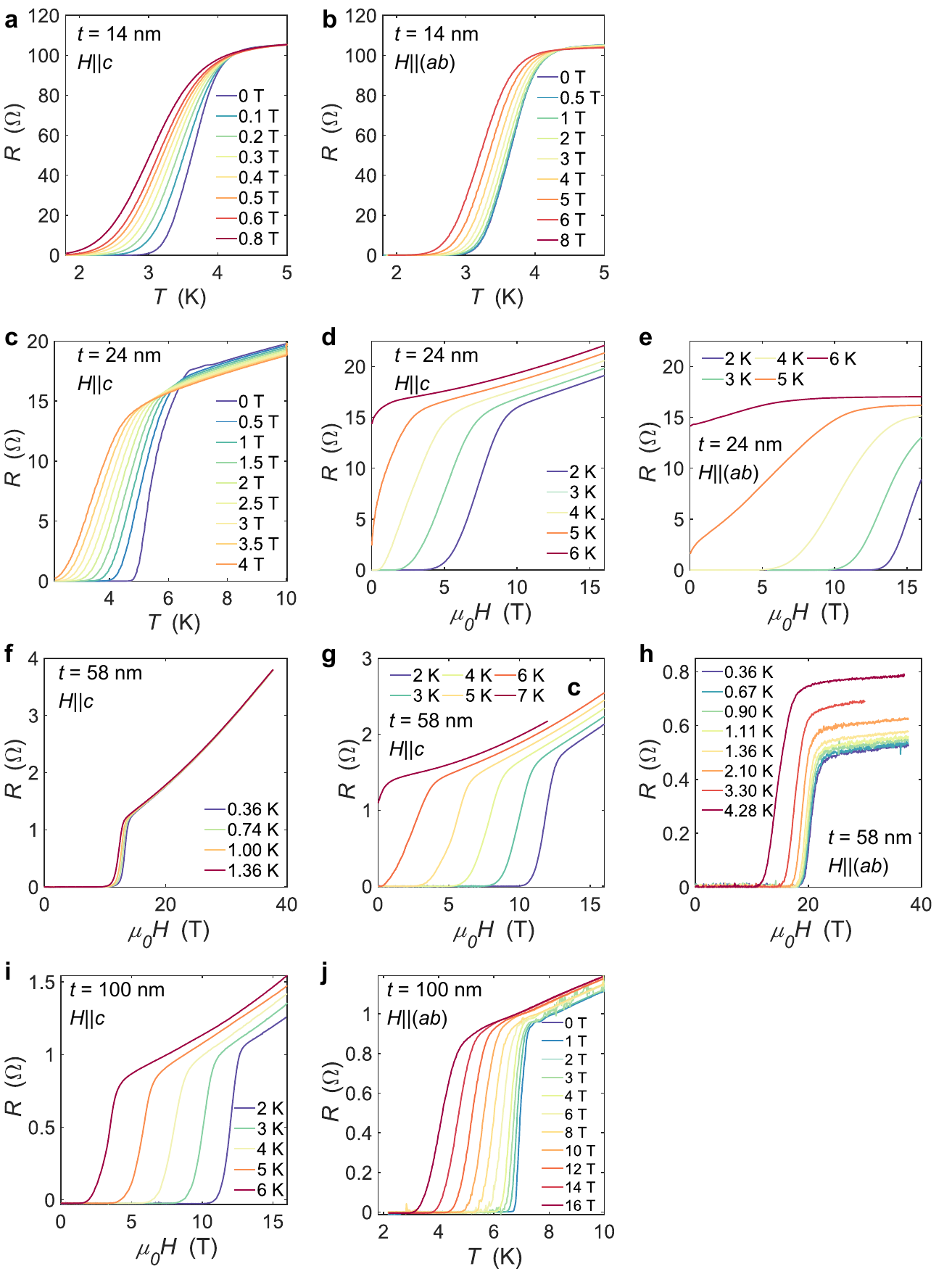}
 \caption{Temperature dependence and angular magnetic field dependence of the resistance of FeSe devices with thickness 14-100 nm.}
 \label{fig:Hc2_Phase_Raw}
\end{figure*}

\begin{figure*}[ht]
 \centering
 	\includegraphics[trim={0cm 0cm 0cm 0cm}, width=0.9\linewidth,clip=true]{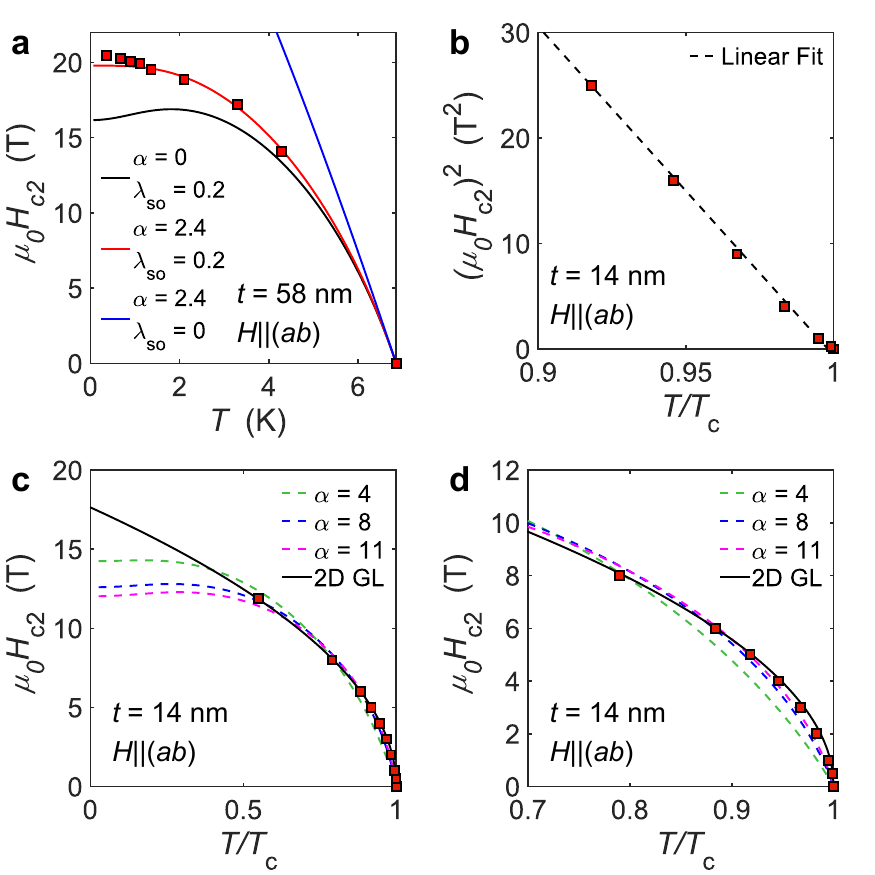}
 \caption{a) Temperature dependence of the upper critical field of a $t$=58~nm sample with an applied field $||(ab)$. The solid lines are fits to WHH theory with varied $\alpha$ and $\lambda_{\rm{so}}$.
b) The upper critical field squared as a function of reduced temperature for a $t$=14~nm device. The dashed line is a linear fit, indicating that $H_{\rm{c2}}||(ab)$ follows a square root dependence near $T_{\rm{c}}$
c) The upper critical field as a function of reduced temperature for a $t$=14~nm device. The dashed lines indcate fits to WHH theory with varied $\alpha$ with $\lambda_{\rm{so}}$ kept at a value of 0.35. The solid line is a fit to 2D-GL theory.
d) A close of c) near $T_{\rm{c}}$.
}
 \label{fig:WHH_Fitting_Example}
\end{figure*}

\begin{figure*}[ht]
 \centering
  	\includegraphics[trim={0cm 0cm 0cm 0cm}, width=0.9\linewidth,clip=true]{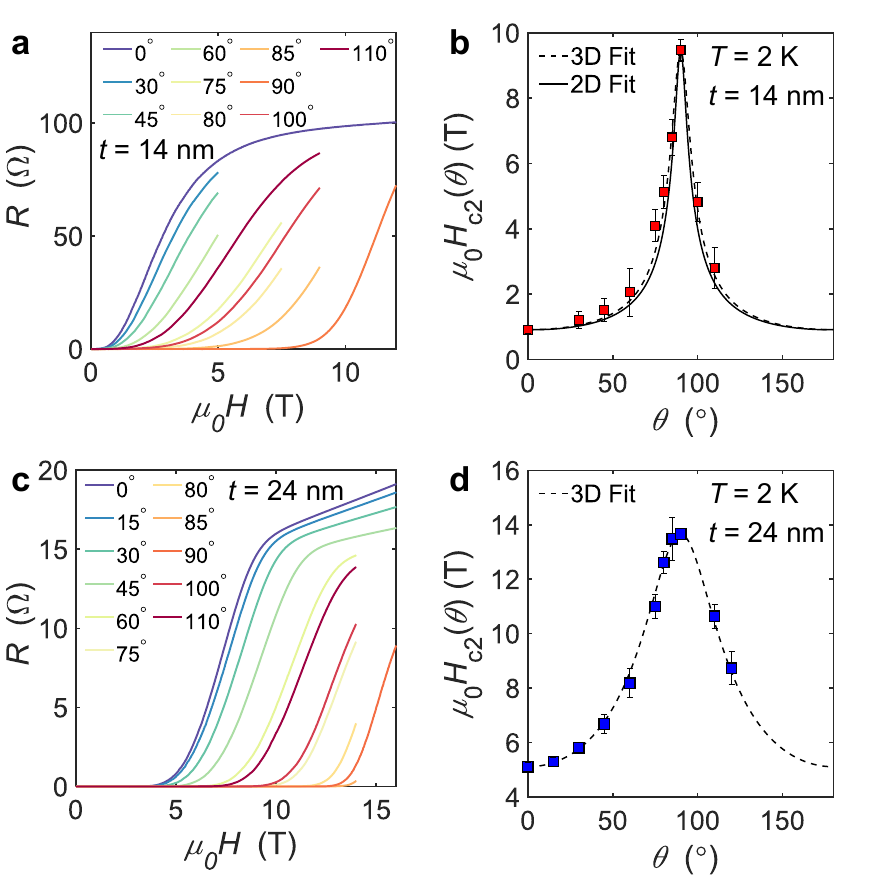}
 \caption{Angular dependence of the resistance transition and upper critical field $H_{\rm c2}(\theta)$ at 2~K for a 14~nm (a) and 24~nm (c) sample. In b,d) we plot the offset value of $H_{\rm c2}$ defined as a linear fit between the superconducting state and the phase transition. Dashed lines are fits to the GL theory for an anisotropic type-II superconductor. The solid line in b) is a fit to the 2D-Tinkham expression for $H_{\rm c2}(\theta)$.}
 \label{fig:Angular_Dependence_Raw}
\end{figure*}

\begin{figure*}[ht]
 \centering
   	\includegraphics[trim={0cm 0cm 0cm 0cm}, width=0.9\linewidth,clip=true]{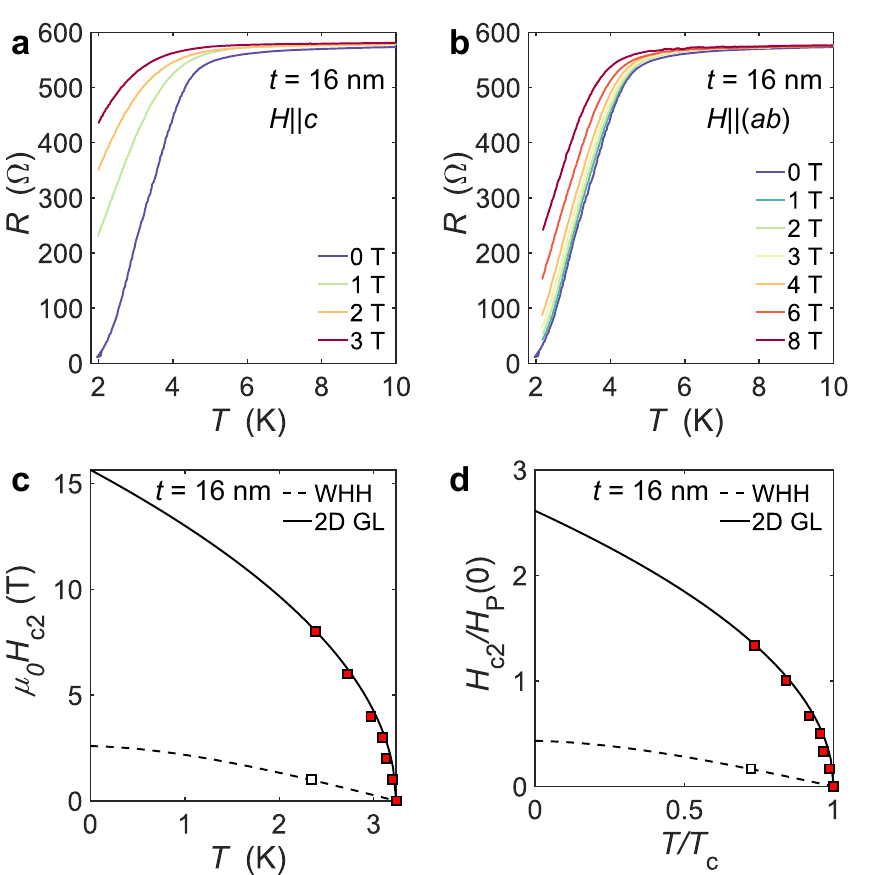}
 \caption{a,b) Superconducting transition of a $t$=16 nm FeSe thin flake device as a function of magnetic field applied parallel and perpendicular to the crystallographic $c$-axis.
	c) Temperature dependence of $T_{\rm c}$ as a function of magnetic field applied parallel and perpendicular to the crystallographic $c$ axis.
$T_{\rm c}$ were determined by the resistive transition midpoint. Dashed lines represent fits to the WHH orbital pair-breaking expectation for $H_{\rm{\rm c2}}$($T$) for the case of $H \parallel c$ and 2D-GL theory for $H \parallel$($ab$).
	d) Reduced temperature dependence $T/T_{\rm c}$ as a function of the ratio between
in-plane upper critical field and the Pauli limited critical field $H_{\rm c2}/H_{\rm{P}}$.}
 \label{fig:FeSeLF9}
\end{figure*}

\begin{figure*}[ht]
 \centering
    	\includegraphics[trim={0cm 0cm 0cm 0cm}, width=0.9\linewidth,clip=true]{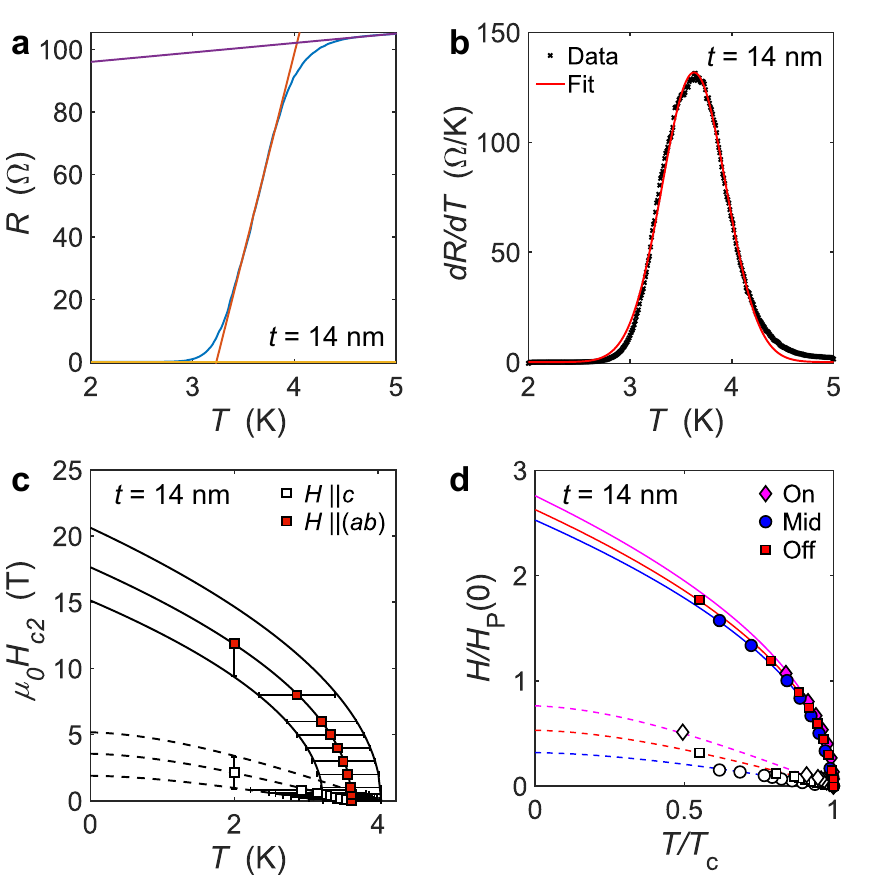}
 \caption{a) The superconducting transition temperature offset value $T^{off}_{\rm c}$ is defined as the intersect of a linear fit between the superconducting state and the resistance transition. The onset value $T^{\rm on}_{\rm c}$ is defined as the intersect of a linear between the resistance transition and the normal state resistance.
	b) The midpoint value $T^{\rm mid}_{\rm c}$ is defined as the maximum value of a fit to the derivative of the resistance transition.
 c) Temperature dependence of $H_{\rm c2}$ as a function of magnetic field parallel and perpendicular to the crystallography $c$ axis of the 14~nm flake. The error bars are the onset and offset values of the superconducting transition as detailed in a) and b).
 d) The phase diagram as in c) but
 in reduced units of reduced temperature  $T/T_{\rm c}$ and magnetic field, $H/H_{\rm P}$, where $H_{\rm P}$ is the Pauli limited critical field.}
 \label{fig:Anisotropy_Comparison}
\end{figure*}

\begin{figure*}
\centering
    	\includegraphics[trim={0cm 0cm 0cm 0cm}, width=0.9\linewidth,clip=true]{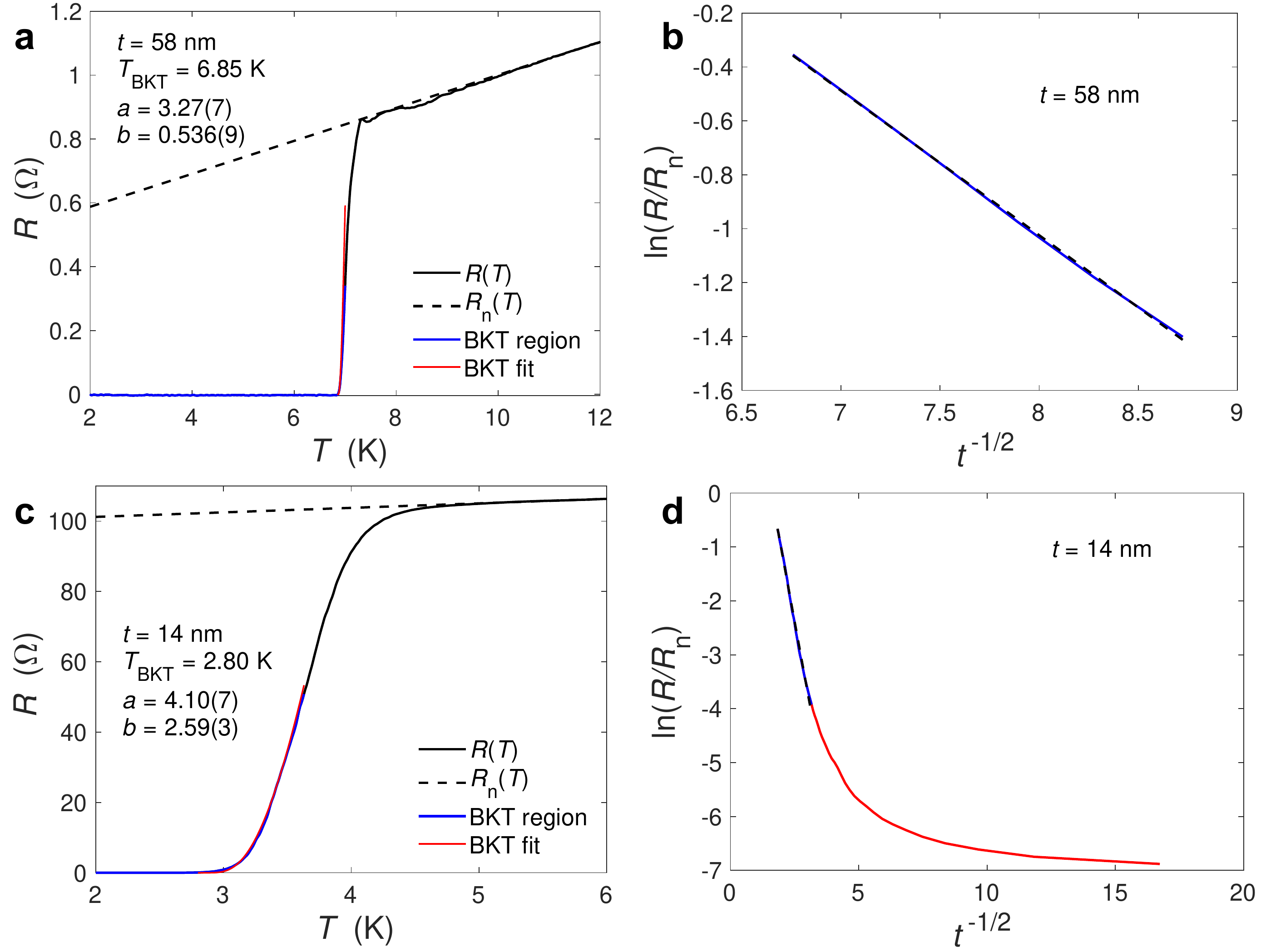}
 \caption{Resistance against temperature in the vicinity of the superconducting transition for two samples with thickness
 $t=14$ in a) and $58\,$nm in c).  The low temperature resistive part
 is fitted to $R = R_{\rm{n}}e^{a-bt^{-1/2}}$, where $R_{\rm{n}}$ is the normal state resistance \cite{Schneider2014}.
  The dashed lines are fits to the high-temperature regime above $T_{\rm{c}}$,   $a$ and $b$ are constants and $t=(T-T_{\rm{BKT}})/T_{\rm{BKT}}$
  is the reduced temperature.
  b), d) The dependence $\rm{ln}(R/R_{\rm{n}})$ against $t^{-1/2}$ for the two different samples from a) and c).
   A linear dependence is evidence that the lowest temperature finite resistance region can be described in terms of BKT effects.}
 \label{fig:BKT_transition_fit}
\end{figure*}

\end{document}